\documentclass[twocolumn,jabbrv]{aastex631}

\usepackage{jabbrv}
\usepackage{tikz}
\UseRawInputEncoding
\usepackage{mathtools}
\usepackage{xcolor}

\usepackage{natbib}
\definecolor{BLUE}{rgb}{0.2,0.2,1}

\graphicspath{{./}{figures/}}

\usepackage{mathrsfs}
\usepackage{amssymb}
\usepackage{amsmath}
\usepackage{graphicx}
\usepackage[normalem]{ulem}
\usepackage{bm}
\usepackage{longtable}
\usepackage{slashed}
\renewcommand\theequation{\arabic{equation}}
\usepackage[T1]{fontenc}
\usepackage{times}
\usepackage{fouriernc}
\usepackage{array}

\usepackage{hyperref}
\hypersetup{colorlinks,linkcolor={cyan},citecolor={magenta},urlcolor={magenta}}  

\renewcommand\sout{\bgroup \color{red} \ULdepth=-.5ex \ULset}

\renewcommand{\v}[1]{\textbf{#1}}
\renewcommand{\rm}[1]{\textrm{#1}}
\renewcommand{\d}{\mathrm{d}}

\renewcommand{\baselinestretch}{0.98}

\makeatletter
\let\frontmatter@title@above=\relax
\makeatother

\renewcommand\thesection{\arabic{section}}

\let\oldbibliography\thebibliography
\renewcommand{\thebibliography}[1]{
  \oldbibliography{#1}
  \setlength{\itemsep}{1pt}
  \setlength{\baselineskip}{10.pt}
  \setlength{\lineskiplimit}{-\maxdimen}
}
\usepackage{jabbrv}

\usepackage{subfigure,dcolumn,jabbrv}
\usepackage{graphicx}
\usepackage{dcolumn}
\usepackage{bm}
\usepackage{CJK}
\usepackage{url}
\usepackage{color}
\usepackage{booktabs}
\usepackage{natbib}
\usepackage{hyperref}
\usepackage{cleveref}
\renewcommand{\thefootnote}{$\star$}

\begin{document}

\title{How Neutron Star Radii Encode the Dense-Matter Equation of State and Hadron-Quark Transition}
\author{Bao-An Li}
\affiliation{Department of Physics and Astronomy,
East Texas A\&M University, Commerce, TX 75429-3011, USA}
\author{Xavier Grundler}
\affiliation{Department of Physics and Astronomy,
East Texas A\&M University, Commerce, TX 75429-3011, USA}

\correspondingauthor{Bao-An Li}
\email{Bao-An.Li@etamu.edu}

\date{\today}

\fontdimen2\font=2.pt

\begin{abstract}
We investigate how future high-precision neutron star (NS) radius measurements
encode microscopic information about the dense-matter equation of state
(EOS), focusing on a possible first-order hadron--quark phase transition
and the resulting mass--radius topology. Within a Bayesian framework using
meta-model EOSs with nine microscopic parameters, we analyze mock radius 
measurements $R_{1.4}=11.9\pm\sigma_R$ km with
$\sigma_R=0.9$ and $0.1$ km for canonical NSs. 
Rather than relying only on marginalized posterior probability distributions, we introduce inverse EOS--radius mappings by studying the conditional mean of each EOS parameter as a function of the model-predicted canonical radius $R_{1.4}$. Their slope measures the sensitivity of an EOS parameter to the radius, while their curvature determines the leading precision dependence of its posterior mean through the Jensen expansion. Resolving the mappings into four mass--radius
topologies, Connected, Disconnected, Both, and No-Quark-Matter, reveals a
clear hierarchy of information. The symmetry-energy parameters $L$ (slope) and
$K_{\rm sym}$ (curvature) are strongly encoded in $R_{1.4}$ and their posterior means
shift appreciably with improved radius precision, whereas the higher-order
hadronic parameters show stronger topology
dependence. Among the transition parameters, the transition density
$\rho_t$ is the most strongly encoded in $R_{1.4}$, while the energy-density
jump and quark-matter sound speed are more strongly associated with the
topology of the full mass--radius sequence. Since the different topologies
have strongly overlapping $R_{1.4}$ distributions, even precise radius
measurements cannot by themselves identify the topology or uniquely
determine the high-density transition properties. These results provide a
parameter-dependent hierarchy for assessing the scientific return of
future high-precision radius measurements and complementary probes of high-density
NS EOS.
\end{abstract}
\section{Introduction}
\label{sec:intro}

One of the central goals of modern neutron star (NS) physics is to infer
the microscopic properties of supradense strongly interacting matter from
macroscopic astronomical observations. Recent advances in X-ray 
observations by NICER and gravitational-wave detections by the
LIGO/Virgo collaboration have already improved our knowledge about 
the equation of state (EOS) of neutron-rich matter over a wide range of densities, for a recent review, see, e.g., refs. \citep{Oertel:2016bki,Baiotti:2019sew,Li:2021thg,Lattimer:2021emm,Sedrakian:2022ata,Chatziioannou:2024jsr}. With the next generation of X-ray timing missions \citep{AngLi25,Cruise} and third-generation gravitational-wave
detectors \citep{Hild:2009ns,Sathyaprakash:2012jk,LIGOScientific:2020zkf,Evans:2021gyd}, the uncertainty in NS radius measurements is
expected to improve by almost an order of magnitude over the coming
decade \citep{Chatziioannou:2021tdi,Pacilio:2021jmq,Finstad:2022oni,Bandopadhyay:2024zrr,Walker:2024loo}. This prospect naturally raises an important question:
\emph{How is microscopic dense-matter physics encoded into measurable
NS radii?}

Bayesian inference has become the standard framework for extracting EOS
information from NS observations. Numerous studies have
shown how current and future measurements constrain supradense neutron-rich nuclear matter
properties and possible hadron--quark phase transitions as well as possible appearance of various non-nucleonic degrees of freedom 
by determining posterior probability distribution functions (PDFs) of EOS
parameters. Although these PDFs quantify the
resulting parameter constraints, they provide limited insight into the
underlying physical mechanism through which different microscopic EOS
parameters are encoded into observable stellar radii. In particular,
it remains unclear why improving the radius precision dramatically
changes the inferred values of some EOS parameters while leaving
others almost unaffected.

This question becomes especially important when considering
first-order hadron--quark phase transitions. Such transitions may
produce twin NSs, namely two stars with identical
gravitational masses but different radii and internal compositions \citep{gerlach_1968,Kampfer:1981yr,Glendenning:1998ag,Schertler_2000}.
Whether the resulting mass--radius sequence remains connected or
develops a disconnected hybrid-star branch depends sensitively on the
high-density behavior of the EOS \citep{Alford:2013aca,zdunik2013,Benic:2014jia,Alvarez_Castillo_2016,Kaltenborn:2017hus,Christian_2018,Alvarez-Castillo:2018pve,Tsa23,Gorda_2023,Carlomagno:2023nrc,Jimenez:2024hib,Albino:2024ymc,Li_2024,Chanlaridis:2024rov,Veselsky:2024bnf,Christian_2025,Pal_2025,Laskos_Patkos_2025,huang2025,Haque26,Haque:2026dre}. 
Consequently, future high-precision
radius measurements offer not only the possibility of discovering twin
stars but also of probing the microscopic physics responsible for the
phase transition \citep{Li:2025tku,Li:2026hnz}. Understanding exactly how this information is encoded
into observable radii is therefore essential for interpreting future
observations.

Our recent Bayesian studies have investigated how future radius
measurements improve constraints on hybrid-star EOS parameters \citep{Li:2024imk,Li:2025tku,Li:2026zuq}.
Using mock radius measurements with progressively smaller
uncertainties, we demonstrated that the hadron--quark transition
density is significantly better constrained than the energy density jump and
the quark-matter sound speed. However, the physical origin of this
behavior remained unclear because all accepted EOSs were analyzed as a
single ensemble without distinguishing their mass--radius topologies.
Likewise, our recent study \citep{Li:2026ult} of Jensen corrections \citep{Jen,Ber,Ebook,infobook} demonstrated that
nonlinear inverse mappings between observables and EOS parameters
naturally produce systematic shifts in Bayesian posterior means as the
measurement precision improves. The connection between these nonlinear
mappings and the topology of hybrid-star sequences, however, has not
been investigated.

In this work we address these open questions by introducing inverse
EOS--radius mappings as a physical interpretation of Bayesian
inference. Rather than considering only the posterior PDFs of EOS parameters, we investigate how the posterior mean
of each microscopic EOS parameter varies with the inferred NS
radius after marginalizing over all remaining parameters. These
inverse mappings reveal how efficiently different microscopic
properties are encoded into observable radii through the Tolman--Oppenheimer--Volkoff (TOV)
equations \citep{tolman1934effect,oppenheimer1939massive}. They further demonstrate that the
first and second derivatives of the mappings quantify two distinct
aspects of the inference process: the slope measures the sensitivity of the inferred EOS parameter to the observed radius variation, whereas the curvature determines the sensitivity of the
posterior mean to improvements in the observational precision through
the Jensen expansion \citep{Li:2026ult}.

\begin{figure*}
\centering
\resizebox{\textwidth}{!}{%
\begin{tikzpicture}
\node at (2.5,4.5) {No-Quark-Matter};
\draw[ultra thick, ->] (0,0) -- (5,0) node[below] {$R$};
\draw[ultra thick, ->] (0,0) -- (0,4.5) node[above] {$M$};
\draw[ultra thick] (4,0.25) to[out=160,in=180+100] (2,3.25);
\draw[thick,dashed,red] (2,3.25) to[out=180+100,in=180+190] (0.5,2.25);
\draw[ultra thick] (4.5,0.5) to[out=140,in=180+90] (3.25,3) to[out=90,in=180+180] (2.5,4);
\draw[thick,dashed] (2.5,4) to[out=180,in=180+225] (1.5,3.5);
\end{tikzpicture}%
\hspace{0.3cm}%
\begin{tikzpicture}
\node at (2.5,4.5) {Both connected \& disconnected};
\draw[ultra thick, ->] (0,0) -- (5,0) node[below] {$R$};
\draw[ultra thick, ->] (0,0) -- (0,4.5) node[above] {$M$};
\draw[ultra thick] (4.5,0.5) to[out=160,in=180+120] (3.25,2);
\draw[ultra thick,red] (3.25,2) to[out=120,in=180+180] (2.5,2.75);
\draw[thick,red,dashed] (2.5,2.75) to[out=180,in=180+180] (2,2.5);
\draw[ultra thick,red] (2,2.5) to[out=180,in=180+180] (1.25,3.5);
\draw[thick,red,dashed] (1.25,3.5) to[out=180,in=180+225] (0.5,3);
\end{tikzpicture}%
\hspace{0.3cm}%
\begin{tikzpicture}
\node at (2.5,4.5) {Connected};
\draw[ultra thick, ->] (0,0) -- (5,0) node[below] {$R$};
\draw[ultra thick, ->] (0,0) -- (0,4.5) node[above] {$M$};
\draw[ultra thick] (4.5,0.5) to[out=160,in=180+120] (3,2.5);
\draw[ultra thick,red] (3,2.5) to[out=120,in=180+180] (2,4);
\draw[thick,red,dashed] (2,4) to[out=180,in=180+245] (1.25,3.25);
\end{tikzpicture}%
\hspace{0.3cm}%
\begin{tikzpicture}
\node at (2.5,4.5) {Disconnected};
\draw[ultra thick, ->] (0,0) -- (5,0) node[below] {$R$};
\draw[ultra thick, ->] (0,0) -- (0,4.5) node[above] {$M$};
\draw[ultra thick] (4.5,0.5) to[out=160,in=180+120] (3,2.5);
\draw[thick,red,dashed] (3,2.5) to[out=180+120,in=180+180] (2.5,1.5);
\draw[ultra thick,red] (2.5,1.5) to[out=180,in=180+180] (1.25,3.5);
\draw[thick,red,dashed] (1.25,3.5) to[out=180,in=180+245] (0.5,2.75);
\end{tikzpicture}%
}
\caption{Modified from similar figures in Refs.~\cite{Alford:2013aca,Alford:2015gna,Zhang:2023wqj,Grundler:2025mcz}, they show an exaggerated mass--radius sequence for each topology category. The change from black to red marks the appearance of QM in the core. Dashed lines represent unstable configurations.}
\label{skt}
\end{figure*}
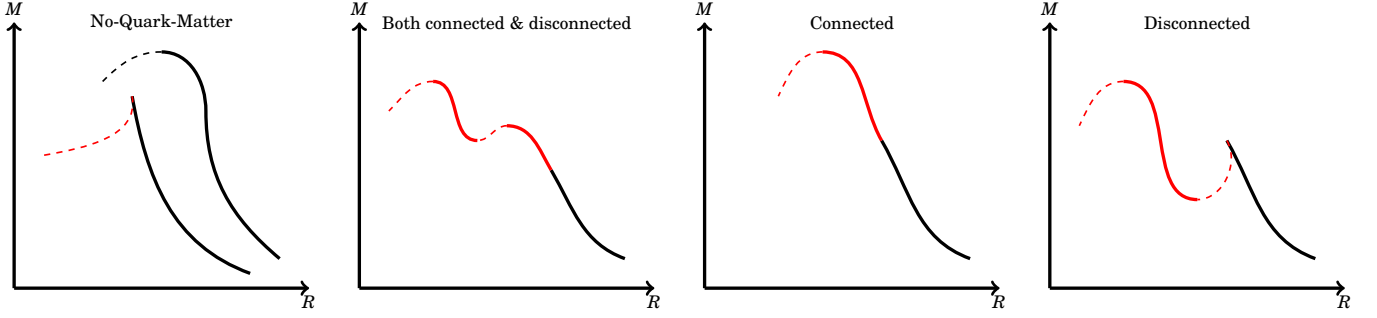

Using a meta-model EOS \citep{zhang2018combined,Zhang:2018bwq,Zhang:2020zsc} encapsulating a first-order hadron--quark phase
transition \citep{xie2021bayesian,Zhang:2023wqj}, we perform Bayesian analyses employing mock measurements
of the canonical NS radius
$R_{1.4}=11.9\pm\sigma_R$ (km) with $\sigma_R=0.9$ and 0.1 representing the current and expected future precision, respectively.
The accepted EOSs are subsequently classified according to the
topology of their mass--radius sequences into Connected (hybrid star branch connected directly to the hadronic branch),
Disconnected (hybrid and hadronic branches are separated by an unstable region), Both (both connected and disconnected hybrid branches coexist), and No-Quark-Matter (purely hadronic stars consisting of neutrons, protons, electrons and muons) categories \citep{Alford:2013aca}. 
These topologies are sketched in Fig. \ref{skt} for ease of the following discussions.

The topology-resolved inverse mappings reveal a clear hierarchy in the
observability of microscopic EOS parameters from NS radii. 
We show that the symmetry-energy parameters $L$ and $K_{\rm sym}$ governing the pressure in canonical NSs around $2\rho_0$ 
are more directly encoded in $R_{1.4}$, whereas the higher-density
hadronic parameters and hadron--quark transition properties
are more strongly associated with the topology of the mass--radius
sequence. Future high-precision radius measurements therefore constrain different microscopic aspects of the NS EOS and hadron--quark transition with markedly different efficiencies.

The remainder of this paper is organized as follows.
Section~II briefly summarizes the Bayesian framework, the EOS model, the approach for obtaining the EOS-radius inverse mapping and its
connection with the Jensen expansion. 
Sections~III and IV discuss the inverse mappings and topology dependence of hadronic and
quark-matter EOS parameters, respectively. Section V is devoted to analyzing the posterior PDFs as well as their dependence on the precision of NS radius measurement and topology of the mass-radius sequence. The resulting observability hierarchy and its implications for future
high-precision NS observations are summarized in the final
section.

\section{Bayesian Framework and Inverse EOS--Radius Mappings}
\label{sec:framework}

Our Bayesian framework follows closely that developed in
Refs.~\cite{xie2019bayesian,xie2020bayesian,Xie:2024mxu,Grundler:2025mcz}, and only the essential
ingredients are summarized here. The objective of the present work is
not to introduce a new Bayesian methodology, but rather to understand
how microscopic properties of dense matter are encoded into observable
NS radii and why different EOS parameters respond
differently to improvements in observational precision.

\subsection{Meta-model NS EOS and priors of its parameters}
\begin{table}[htbp]
\centering
\caption{Prior ranges of the EOS parameters in units of MeV and the critical density $\rho_t/\rho_0$ for hadron-quark phase transition.}\label{tab-prior}
\begin{tabular}{lcc}
\hline\hline
Parameters & Lower limit & Upper limit \\
\hline
$K_0$ & 220 & 260 \\
$J_0$ & -400 & 400 \\
$K_{\mathrm{sym}}$ & -400 & 100 \\
$J_{\mathrm{sym}}$ & -200 & 800 \\
$L$ & 30 & 90 \\
$E_{\mathrm{sym}}(\rho_0)$ & 28.5 & 34.9 \\
$\Delta\epsilon/\epsilon_t$ & 0.2 & 1.0 \\
$c^2_{\rm s}/c^2$ & 0.0 & 1.0 \\
$\rho_t/\rho_0$ & 3.0 & 6.0 \\
\hline
\end{tabular}
\end{table}

The NS EOS is described using the meta-model which combines a flexible
parameterization of the hadronic EOS (for a recent review, see, e.g., refs.~\citep{Li:2024imk,Li:2025tku}) with a first-order
hadron--quark phase transition to quark matter with a constant speed of sound (CSS) \citep{Alford:2013aca,zdunik2013,Alford:2015dpa}. For ease of the following discussions, we briefly recall here a few relevant definitions and terminologies involved in the meta-model EOS for NS cores. In the CSS model, the energy density can be written as \citep{Alford:2015gna}
\begin{equation}
\varepsilon(p)= \begin{cases}\varepsilon_{\mathrm{HM}}(p) & \rho<\rho_{t} \\ \varepsilon_{\mathrm{HM}}\left(p_{t}\right)+\Delta \varepsilon+C_{\mathrm{s}}^{-2}\left(p-p_{t}\right) & \rho>\rho_{t}\end{cases}
\end{equation}
where $\varepsilon_{\mathrm{HM}}(p)$ is the energy density of hadronic matter (HM) at pressure $p$, $p_t$ is the pressure at the transition density $\rho_t$, $\Delta \varepsilon$ describes the strength (latent heat) of the phase transition, and the speed of sound squared $C_{\rm{s}}^2$ quantifies the stiffness of quark matter.

Following the classification scheme of Alford {\it et al.}~
\cite{Alford:2013aca,Alford:2015gna}, the resulting stellar sequences
are classified into four topological categories: Connected,
Disconnected, Both, and No-Quark-Matter. These categories represent
qualitatively distinct realizations of the hadron--quark phase
transition and form the basis of the topology-resolved analyses
presented below.

We note that the unstable configurations (dashed branches) in
Fig.~\ref{skt} are identified using the conventional turning-point
criterion \cite{stableMethod, Alford_2017, Essick:2024olf}, which corresponds to the rapid phase-conversion limit. For
a sharp phase boundary, however, the stability of configurations can
depend on the timescale of phase conversion. In the slow-conversion
limit, configurations with $dM/d\rho_c<0$ can remain dynamically
stable, as demonstrated in Ref.~\citep{Pereira:2017rmp}. The stability
structure can be further modified in the presence of sequential phase
transitions, as discussed in Ref.~\citep{Rau:2022ofy}. In particular,
sequential transitions between distinct quark phases can generate
additional stable hybrid branches, including hybrid twins and even
mass triplets \citep{Alford:2017qgh}. The present analysis is
restricted to a single hadron--quark transition and adopts the
conventional rapid-conversion stability criterion. Accordingly, the
topological classifications and stability properties discussed below
should be understood within these assumptions.

The hadronic EOS $\varepsilon_{\mathrm{HM}}(p)$ is characterized by the empirical parameters of symmetric nuclear matter and the symmetry
energy. More specifically, the binding energy $E_0(\rho)$ per nucleon in symmetric nuclear matter (SNM) and nuclear symmetry energy $E_{\rm{sym}}(\rho)$ are parameterized as
\begin{eqnarray}\label{E0para}
  E_{0}(\rho)&=&E_0(\rho_0)+\frac{K_0}{2}(\frac{\rho-\rho_0}{3\rho_0})^2+\frac{J_0}{6}(\frac{\rho-\rho_0}{3\rho_0})^3,\\
  E_{\rm{sym}}(\rho)&=&E_{\rm{sym}}(\rho_0)+L(\frac{\rho-\rho_0}{3\rho_0})+\frac{K_{\rm{sym}}}{2}(\frac{\rho-\rho_0}{3\rho_0})^2\nonumber\\
  &+&\frac{J_{\rm{sym}}}{6}(\frac{\rho-\rho_0}{3\rho_0})^3\label{Esympara},
\end{eqnarray}
where $E_0(\rho_0)=-16$ MeV at the SNM saturation density $\rho_0=0.16/\rm{fm}^3$. The parameters $K_0$ and $J_0$ measure the stiffness of 
SNM EOS. Expanding around the saturation density $\rho_0$, the parameters $E_{\rm{sym}}(\rho_0)$, $L$, $K_{\rm{sym}}$, and $J_{\rm{sym}}$ characterize the magnitude, slope, curvature, and skewness of nuclear symmetry energy $E_{\rm{sym}}(\rho)$ if the latter is known a priori, respectively. In our Bayesian analyses, they are just parameters generated randomly within the prior ranges listed in Table \ref{tab-prior}. Nevertheless, it is worth noting that the skewness parameters $J_0$ and $J_{\rm{sym}}$ characterize the stiffness of the generated $E_0(\rho)$ and $E_{\rm{sym}}(\rho)$ around $(3-4)\rho_0$ \citep{Xie:2020kta}, while $K_{\rm{sym}}$ and $L$ characterize the stiffness of $E_{\rm{sym}}(\rho)$ around $(1-3)\rho_0$ and $\rho_0$, respectively. 

In the present work, we adopt the prior range 
$
3\leq\rho_t/\rho_0\leq6,
$
for the hadron-quark transition density and construct the $R_{1.4}$ inverse mappings using the first (hadronic)
branch. Given this prior, all accepted EOSs support a $1.4M_\odot$ star
on the first branch. This choice deserves special discussion, as some of our conclusions depend on it, and many studies in the literature use a lower $\rho_t$-boundary that leads to quark matter formation in lighter NSs; see, e.g., Ref. \citep{Li_2024} and a brief review in Section 2.4 of Ref. \citep{Li:2025tku}. 
Indeed, a prior extending down to $\rho_t\sim\rho_0$ would be a
straightforward numerical modification, as we have done previously in Refs. \citep{Li:2025tku,Li:2026hnz}. However, we do \emph{not} regard
such a change as physically innocuous. The lower limit of the transition
density is itself a physics input. It should not be selected solely by
making the prior as broad as possible or by following the location of a
posterior peak obtained from NS data alone. This issue was discussed explicitly in Section~2.4 of our earlier paper \cite{Li:2025tku}. In that work we deliberately
compared the widely used $(1$--$6)\rho_0$ prior (case A) with the
BES/RHIC-informed $(3$--$6)\rho_0$ prior (case B). With the broader prior,
the NS Bayesian analysis produced a prominent posterior peak
near $(1.7$--$2.0)\rho_0$. However, our analysis showed that this feature is
statistically sufficient to accommodate the presently available NS data but is not
physically required once the BES/RHIC experimental indications \citep{STAR3,STAR2,STAR:2025owm} are taken
into account. The corresponding posterior structure in the $(3$--$6)\rho_0$
analysis is associated with the higher-density feature already present in
the broader-prior calculation. The physical motivation for the lower bound of $3\rho_0$ comes from the
recent BES (Beam Energy Scan) program at RHIC by the STAR Collaboration. STAR measurements
indicate that QGP signatures disappear in Au+Au collisions at
$\sqrt{s_{NN}}=3$ GeV \citep{STAR3}, where collective flow observables are reproduced
by hadronic transport calculations, while signatures of partonic
collectivity reemerge at $\sqrt{s_{NN}}=4.5$ GeV \cite{STAR:2025owm}. In an updated ART \citep{ART}
transport calculation \citep{Yong22}, the maximum central baryon density reached at
$\sqrt{s_{NN}}=3$ GeV is about $3.6\rho_0$. These observations provide
strong circumstantial evidence that the hadron--quark transition density
in hot dense matter should lie above approximately $3.6\rho_0$.
We do not assume a one-to-one correspondence between the matter created
in heavy-ion collisions and NS matter. The systems differ
in temperature, isospin asymmetry, equilibrium conditions, and composition.
Rather, the argument uses the fact that both systems probe different
regions of the same QCD phase diagram. The first-order hadron-quark transition boundary connects
these regions, and most theoretical studies indicate that the transition
line generally moves to higher density as the temperature decreases,
although the precise isospin dependence remains model-dependent. We
therefore regard the BES/RHIC information on the lower boundary of $\rho_t$ as an important nuclear-physics
constraint on the physically admissible prior, not as a replacement for NS observations.

This point is important for interpreting Bayesian inference. A posterior
probability obtained from NS data alone cannot by itself decide
whether a statistically preferred region of parameter space is physically
realistic when relevant information from terrestrial experiments has not been included.
In this sense, comprehensive statistical inference should incorporate, rather than
replace, physics constraints established by independent observations and experiments.
Our choice of $3\leq\rho_t/\rho_0\leq6$ is therefore a deliberate
terrestrial experiments-informed prior rather than an arbitrary restriction imposed for
computational convenience. Nevertheless, it is important to emphasize that some of our conclusions are conditioned on this choice. Moreover, a broader analysis incorporating alternative, independently
justified transition-density priors would be valuable in future work.

\subsection{Bayesian framework}
For each EOS parameter set $\boldsymbol{\theta}$, the TOV equations are solved to obtain the mass-radius sequence. In our Bayesian analyses, the total likelihood function has three factors
\begin{equation}\label{Likelihood}
  {\cal L}(D|{\cal \boldsymbol{\theta}}) = {\cal L}_{\rm{filter}}(D|{\cal \boldsymbol{\theta}})\times {\cal L}_{\rm{mass,max}}(D|{\cal \boldsymbol{\theta}})\times {\cal L}_\mathrm{R_{1.4}}(D|{\cal \boldsymbol{\theta}})
\end{equation}
where ${\cal L}_{\rm{filter}}(D|{\cal \boldsymbol{\theta}})$ and ${\cal L}_{\rm{mass,max}}(D|{\cal \boldsymbol{\theta}})$ indicate that the generated EOSs must satisfy the following conditions: (i) The crust-core transition pressure remains positive; (ii) The thermodynamic stability condition, $\rm{d}P/\rm{d}\epsilon\geq0$, holds at all densities; (iii) The causality condition is upheld at all densities; (iv) The generated NS EOS should be sufficiently stiff to support NSs at least as massive as 1.97 M$_{\odot}$ (i.e., the minimum M$_{\rm{TOV}}$ which is the maximum mass a given EOS can support). The latter is based on the mass observations of PSR J0348+0432 with $M = 2.01 \pm 0.04$ M$_\odot$ \cite{Antoniadis:2013pzd}. Since this condition was used in deriving constraints on the radius $R_{1.4}$ for canonical NSs from GW170817 by the LIGO/VIRGO Collaboration \cite{abbott2018gw170817}, and we use their result for $R_{1.4}$ as the base of our mock radius data, we adopt here 1.97 M$_{\odot}$ as the minimum M$_{\rm{TOV}}$. For the mock radius data $R_{1.4}^{D}=11.9$ km with uncertainty $\sigma_R$, the likelihood ${\cal L}_\mathrm{R_{1.4}}(D|{\cal \boldsymbol{\theta}})$
is taken to be
\begin{equation}
{\cal L}_\mathrm{R_{1.4}}(D|\boldsymbol{\theta}) \propto
\exp\left[-\frac{\left(R_{1.4}(\boldsymbol{\theta})
-R_{1.4}^{D}\right)^2}{2\sigma_R^2}\right].
\end{equation}
The posterior is then
\begin{equation}
P(\boldsymbol{\theta}|D)\propto
{\cal L}(D|\boldsymbol{\theta})\pi(\boldsymbol{\theta}),
\end{equation}
with $\pi(\boldsymbol{\theta})=1.0$ in the prior ranges given in Table \ref{tab-prior} and zero otherwise. 

\subsection{Inverse EOS--Radius Mappings}

Traditional Bayesian analyses focus on the marginal posterior
probability distributions, $P(\theta_i|D)$, where $\theta_i$ denotes an
EOS parameter and $D$ represents the observational data, subject to the
requirements that the EOS support the observed maximum NS mass,
remain causal and mechanically stable, and satisfy existing constraints
from terrestrial nuclear experiments. These posterior distributions
quantify the uncertainties of the inferred EOS parameters, but they
provide less direct insight into how information from a NS
observable is transmitted to the underlying microscopic EOS.

To expose this connection, we construct inverse mappings by first sorting
all accepted posterior samples according to their model-predicted
canonical NS radius $R_{1.4}$ and grouping them into bins of
width $0.05$ km. Within each radius bin, we calculate the conditional
posterior mean of an EOS parameter $\theta_i$,
\begin{equation}
\Theta_i(R_{1.4})
\equiv
\left\langle\theta_i\mid R_{1.4}\right\rangle
=
\int \theta_i
P(\theta_i\mid R_{1.4})\,d\theta_i.
\label{eq:inverse_mapping}
\end{equation}
The $\Theta_i(R_{1.4})$ as a function of the model-predicted $R_{1.4}$ is referred to as the inverse EOS--radius mapping.
The resulting functions $\Theta_i(R_{1.4})$, shown in Figs.~\ref{HMEvery} and
\ref{QMEvery}, constitute the nearly universal EOS--radius inverse
mappings identified first in our recent work \citep{Li:2026ult}. Importantly,
the conditional expectation in Eq.~(\ref{eq:inverse_mapping}) refers to
a fixed \emph{model-predicted} radius and should therefore be distinguished
from the overall posterior mean of $\theta_i$, which is obtained only
after averaging over the finite-width posterior distribution of predicted $R_{1.4}$.

The near universality of these mappings for different radius measurement uncertainties $\sigma_R$ can be understood analytically. We collect all
factors in the posterior that are independent of the radius likelihood
into
\begin{equation}
Q(\boldsymbol{\theta})
\equiv
{\cal L}_{\rm{filter}}(D|\boldsymbol{\theta})
{\cal L}_{\rm{mass,max}}(D|\boldsymbol{\theta})
\pi(\boldsymbol{\theta}),
\label{eq:Q_theta}
\end{equation}
so that the posterior can be written as
\begin{equation}
P(\boldsymbol{\theta}|D,\sigma_R)
\propto
Q(\boldsymbol{\theta})
\exp\left[
-\frac{
\left(R_{1.4}(\boldsymbol{\theta})-R_{1.4}^{D}\right)^2
}{
2\sigma_R^2
}
\right].
\label{eq:posterior_revised}
\end{equation}

To demonstrate the origin of the universality, consider the conditional
expectation at a fixed model-predicted radius $r$,
\begin{equation}
\begin{aligned}
\left\langle\theta_i\mid r\right\rangle
&=
\frac{
\displaystyle
\int d\boldsymbol{\theta}\,
\theta_i
P(\boldsymbol{\theta}|D,\sigma_R)
\delta\!\left[r-R_{1.4}(\boldsymbol{\theta})\right]
}{
\displaystyle
\int d\boldsymbol{\theta}\,
P(\boldsymbol{\theta}|D,\sigma_R)
\delta\!\left[r-R_{1.4}(\boldsymbol{\theta})\right]
}.
\end{aligned}
\label{eq:conditional_delta}
\end{equation}
Because the delta function imposes
$R_{1.4}(\boldsymbol{\theta})=r$, the radius likelihood contributes the
same multiplicative factor,
\begin{equation}
\exp\left[
-\frac{(r-R_{1.4}^{D})^2}{2\sigma_R^2}
\right],
\end{equation}
to both the numerator and denominator of
Eq.~(\ref{eq:conditional_delta}). This factor therefore cancels, giving
\begin{equation}
\left\langle\theta_i\mid r\right\rangle
=
\frac{
\displaystyle
\int d\boldsymbol{\theta}\,
\theta_i Q(\boldsymbol{\theta})
\delta\!\left[r-R_{1.4}(\boldsymbol{\theta})\right]
}{
\displaystyle
\int d\boldsymbol{\theta}\,
Q(\boldsymbol{\theta})
\delta\!\left[r-R_{1.4}(\boldsymbol{\theta})\right]
}.
\label{eq:universal_mapping}
\end{equation}
Thus, within the present Bayesian setup, the conditional inverse mapping
is independent of the radius-measurement precision $\sigma_R$. This
provides an analytical explanation for the numerical observation that
the inverse mappings obtained with different radius measurement precisions collapse
onto essentially the same curves, apart from finite-bin, edge, and
sampling effects \cite{Li:2026ult}. We thank the referee for pointing out this analytical
interpretation of the numerical universality observed in our Bayesian
samples.

Eq.~(\ref{eq:inverse_mapping}) should be contrasted with the
familiar forward-modeling problem solved by the TOV equations,
$\boldsymbol{\theta}\rightarrow R(M)$, which predicts the NS
mass--radius sequence from a specified EOS. The inverse mapping describes
the opposite inference process, $R\rightarrow\Theta_i(R)$, namely how a
model-predicted NS radius constrains a microscopic EOS
parameter after the remaining uncertainties have been marginalized.
These mappings therefore provide a direct visualization of how
information contained in NS radius measurements is transmitted
back to the underlying dense-matter EOS.

For EOSs exhibiting twin-star topologies, the inverse mappings are
constructed using the $R_{1.4}$ values on the first (hadronic) branch.
Given our chosen prior on the hadron--quark transition density, all
accepted EOSs support a $1.4\,M_\odot$ neutron star on this branch,
whereas only a small fraction support a $1.4\,M_\odot$ star on the
second, disconnected branch. We therefore use the first-branch radius
consistently when constructing the topology-resolved inverse mappings.
Unlike posterior probability distributions, which describe the outcome
of Bayesian inference, the inverse mappings reveal its physical origin.
Their slope characterizes the sensitivity of an inferred EOS parameter
to variations in the radius, while their curvature determines the
leading nonlinear response to a finite-width radius distribution.

\subsection{Nonlinear Filtering: How $\sigma_R$ Changes the Posterior Means of EOS Parameters}

Although the inverse mapping $\Theta_i(R_{1.4})$ is independent of the
measurement precision, the posterior distribution of the model-predicted
radius is not. To make this distinction explicit, we define the marginal
posterior of $R_{1.4}$ as
\begin{equation}
P(R_{1.4}|D,\sigma_R)
=
\int
P(\boldsymbol{\theta}|D,\sigma_R)
\delta\!\left[
R_{1.4}-R_{1.4}(\boldsymbol{\theta})
\right]
d\boldsymbol{\theta}.
\label{eq:radius_marginal}
\end{equation}
Using Eq.~(\ref{eq:posterior_revised}), this becomes
\begin{equation}
P(R_{1.4}|D,\sigma_R)
\propto
Q_R(R_{1.4})
\exp\left[
-\frac{
\left(R_{1.4}-R_{1.4}^{D}\right)^2
}{
2\sigma_R^2
}
\right],
\label{eq:radius_marginal_weighted}
\end{equation}
where
\begin{equation}
Q_R(R_{1.4})
=
\int
Q(\boldsymbol{\theta})
\delta\!\left[
R_{1.4}-R_{1.4}(\boldsymbol{\theta})
\right]
d\boldsymbol{\theta}
\label{eq:QR}
\end{equation}
contains the prior, physical constraints, and all other likelihood
factors after marginalizing over the EOS parameters at fixed
$R_{1.4}$.

Eq.~(\ref{eq:radius_marginal_weighted}) shows that changing
$\sigma_R$ changes the relative weight assigned to different
model-predicted radii. It does not, by itself, imply a change in the
strict mathematical support of the posterior, since the Gaussian
likelihood is nonzero for every finite radius. More importantly, it
shows why the posterior radius need not peak at the central value
$R_{1.4}^{D}$. The posterior is the product of the Gaussian likelihood
and the generally nonuniform EOS-induced density $Q_R(R_{1.4})$.
The location of the posterior mode therefore satisfies
\begin{equation}
\frac{R_{\rm{mode}}-R_{1.4}^{D}}{\sigma_R^2}
=
\left.
\frac{d\ln Q_R}{dR_{1.4}}
\right|_{R_{\rm{mode}}} .
\label{eq:radius_mode}
\end{equation}
Consequently, for finite $\sigma_R$, the posterior mode generally
differs from $R_{1.4}^{D}$ whenever $Q_R$ has a nonzero local slope.
In the limit $\sigma_R\rightarrow0$, the likelihood becomes increasingly
concentrated at $R_{1.4}^{D}$ and the posterior correspondingly approaches
that value.

Eq.~(\ref{eq:radius_marginal_weighted}) also explains why the
posterior need not be symmetric even though the radius likelihood is
Gaussian. The function $Q_R(R_{1.4})$ generally varies across the region
sampled by the likelihood and need not be symmetric about
$R_{1.4}^{D}$. Multiplication by $Q_R$ therefore distorts the Gaussian
likelihood, producing both a displacement of the posterior peak and,
in general, a skewed posterior. As $\sigma_R$ increases, the likelihood
samples a broader region of $Q_R$, making these effects increasingly
important. These features are consistent with the asymmetric posterior
radius distributions observed in our numerical Bayesian analyses
\citep{Li:2024imk,Li:2025tku,Li:2026ult}.

The connection between the precision-independent inverse mapping and
the precision-dependent posterior mean now follows directly from the
law of total expectation:
\begin{equation}
\left\langle\theta_i\right\rangle_{\sigma_R}
=
\int
\Theta_i(R_{1.4})
P(R_{1.4}|D,\sigma_R)\,dR_{1.4}.
\label{eq:nonlinear_filtering}
\end{equation}
This equation makes explicit the two distinct averaging operations in
the inference. First, $\Theta_i(R_{1.4})$ is obtained from the conditional
average of $\theta_i$ at fixed model-predicted radius and defines the
precision-independent inverse mapping. Second, this mapping is averaged
over the $\sigma_R$-dependent posterior radius distribution to obtain
the overall posterior mean. Thus, changing the observational precision
does not change the inverse mapping itself; rather, it changes how
different portions of the same mapping are weighted.

In this sense, the Bayesian inference can be viewed schematically as
[universal inverse mapping
$+$
$\sigma_R$-dependent radius weighting
$\Longrightarrow$
$\sigma_R$-dependent posterior EOS parameter mean].
For sufficiently narrow posterior radius distributions, Eq.~(\ref{eq:nonlinear_filtering}) can then be expanded about the posterior
mean radius, leading naturally to the Jensen expansion discussed below.

\subsection{Connection with the Jensen Expansion}

The physical significance of the inverse mappings becomes particularly
transparent when combined with the Jensen expansion
\citep{Jen,Ber,Ebook,infobook}, which was first employed in our recent
work on Bayesian analyses of neutron-star properties
\citep{Li:2026ult}. In this subsection, we represent the inverse mapping
for a given EOS parameter as a sufficiently smooth function
$\theta_i=f_i(R)$. This analytical representation is introduced here to
derive the Jensen expansion; it should be distinguished from the
numerically extracted conditional mapping
$\Theta_i(R_{1.4})$ obtained directly from the Bayesian posterior
samples using $R_{1.4}$ as the observable. If $f_i(R)$ is sufficiently
smooth over the range sampled by the posterior radius distribution, a
Taylor expansion of $\theta_i$ about the posterior mean radius $\bar R$,
followed by averaging over the radius distribution, gives
\begin{eqnarray}\label{jensen_general}
\langle\theta_i\rangle_{\sigma_R}
&=&
f_i(\bar R)
+f_i'(\bar R)\langle R-\bar R\rangle
+\frac{1}{2}f_i''(\bar R)
\left(\sigma_R^{\rm{post}}\right)^2\nonumber\\
&+&\frac{1}{6}f_i'''(\bar R)
\left\langle(R-\bar R)^3\right\rangle
+\cdots .
\end{eqnarray}
Here,
$\sigma_R^{\rm{post}}\equiv
\sqrt{\langle(R-\bar R)^2\rangle}$
(which is a function of the observational precision $\sigma_R$)
denotes the standard deviation of the posterior distribution
$P(R|D,\sigma_R)$, thereby distinguishing the posterior width from the
observational uncertainty $\sigma_R$ entering the radius likelihood.
The linear term vanishes by definition of $\bar R$, since
$\langle R-\bar R\rangle=0$. For a symmetric posterior, the odd central
moments also vanish, and Eq.~(\ref{jensen_general}) reduces to
\begin{equation}
\begin{aligned}
\langle\theta_i\rangle_{\sigma_R}
&=
f_i(\bar R)
+\frac{1}{2}f_i''(\bar R)
\left(\sigma_R^{\rm{post}}\right)^2
+{\cal O}\!\left[(\sigma_R^{\rm{post}})^4\right].
\end{aligned}
\label{jensen}
\end{equation}
Thus, for a symmetric and sufficiently narrow posterior, the leading
finite-precision correction is determined by the curvature
$f_i''(\bar R)$: a convex mapping produces a positive shift in the
posterior mean, whereas a concave mapping produces a negative shift.
For an asymmetric posterior, as can occur for finite $\sigma_R$ in
Eq.~(\ref{eq:radius_marginal_weighted}), the third and higher central
moments provide additional corrections.

The Jensen expansion is general and applies to any EOS parameter
$\theta_i$ and any observable for which a corresponding inverse mapping
$\theta_i=f_i(R)$ can be established, including $R_{1.4}$, $R_{2.0}$,
and the tidal deformability $\Lambda$. 

Eq.~(\ref{eq:nonlinear_filtering}) demonstrates that two independent geometric
properties of the inverse mapping determine the Bayesian inference.
The first derivative,
$
\frac{d\Theta_i}{dR},
$
measures the sensitivity of the inferred EOS parameter $\theta_i$ to the
observable radius. Parameters with steep inverse mappings are therefore more sensitive to variations in the measured radius. The second derivative,
$
\frac{d^2\Theta_i}{dR^2},
$
governs the Jensen correction and therefore determines how strongly
the inferred posterior mean changes when the observational precision
is improved. Nearly linear mappings produce only small changes in the
posterior mean as $\sigma_R$ decreases, whereas strongly nonlinear
mappings naturally lead to significant systematic shifts. Moreover, convex mappings (with positive second derivative) increase the means, whereas concave mappings decrease them. 

The inverse mappings therefore provide considerably more physical
insight than posterior probability distributions alone. While the
posterior PDFs describe the outcome of Bayesian inference, the
inverse mappings explain why different EOS parameters respond
differently to future high-precision radius measurements. As shown in
the following sections, they reveal a clear hierarchy in the
observability of microscopic EOS parameters and provide a unified
physical interpretation of both the Bayesian posterior distributions
and the topology of the NS mass--radius sequence.

\section{Inverse EOS--Radius Mappings of Hadronic EOS Parameters}
\label{sec:hadronic}

\begin{figure}[thb]
\centering
\vspace{-1.cm}
 \resizebox{0.3\textwidth}{!}{
\includegraphics[width=1.3\textwidth,
 trim=0 80 0 0,
    clip
]{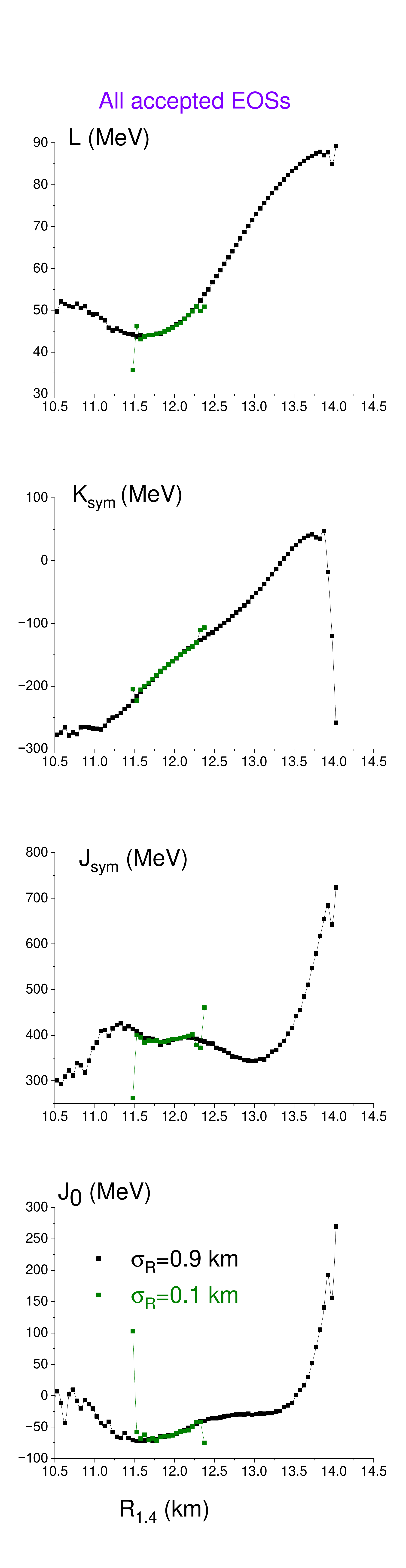}
}
\caption{
Inverse EOS--radius mappings of the four hadronic EOS parameters
$L$, $K_{\rm sym}$, $J_{\rm sym}$, and $J_0$ obtained from all accepted
EOSs. The posterior conditional mean of each parameter is shown as a function of
the inferred canonical radius $R_{1.4}$. The black and green curves
correspond to radius uncertainties $\sigma_R=0.9$ and $0.1$ km,
respectively. 
}
\label{HMEvery}
\end{figure}

We first examine how the microscopic parameters of the hadronic EOS are
encoded in the radius of a canonical NS. Fig.~\ref{HMEvery}
shows the inverse mappings of the four hadronic parameters
$L$, $K_{\rm sym}$, $J_{\rm sym}$, and $J_0$ obtained from all accepted
EOSs. The black and green curves correspond to 
$\sigma_R=0.9$ and $0.1$ km, respectively. The inverse mappings of $K_0$ and $E_{\rm sym}(\rho_0)$
are also examined, but they are largely flat in the relevant radius ranges, indicating their weak dependence on $R_{1.4}$.
They are thus not presented here. These mappings provide a useful first indication of which aspects of the hadronic EOS are
directly encoded in $R_{1.4}$ and which remain largely degenerate with
other EOS parameters.

\subsection{Radius-setting parameters: $L$ and $K_{\rm sym}$}

The most striking feature of Fig.~\ref{HMEvery} is the close similarity
between the inverse mappings of $L$ and $K_{\rm sym}$ obtained with the two very different radius uncertainties.
Both parameters exhibit substantial correlations with $R_{1.4}$ over the radius range populated
by the accepted EOSs. In particular, $K_{\rm sym}$ displays a nearly
monotonic increase with increasing $R_{1.4}$, while $L$ shows a
nonlinear but generally increasing behavior over the central region of
the accepted radius distribution. Thus, a larger canonical radius
requires, on average, a stiffer symmetry-energy sector at
suprasaturation density.

This behavior is consistent with the well-known connection between the
pressure of neutron-rich matter around $(1-2)\rho_0$ and the radius of
a canonical NS \citep{Lattimer:2000nx,LCK}. The parameter $L$ determines the leading
density dependence of the symmetry energy above saturation, while
$K_{\rm sym}$ controls its curvature. Together they provide important
contributions to the pressure in the density range around $2\rho_0$ to which
$R_{1.4}$ is most sensitive. Consequently, different EOSs that
reproduce the same canonical radius are nevertheless required to have
similar effective pressures in this density range. The inverse
mappings therefore show a relatively universal relation between
$R_{1.4}$ and these two symmetry-energy parameters most relevant around $(1-2)\rho_0$.

\begin{figure}[thb]
\centering
\vspace{-0.8cm}
 \resizebox{0.3\textwidth}{!}{
\includegraphics[width=\textwidth,
 trim=0 30 0 0,
    clip
]{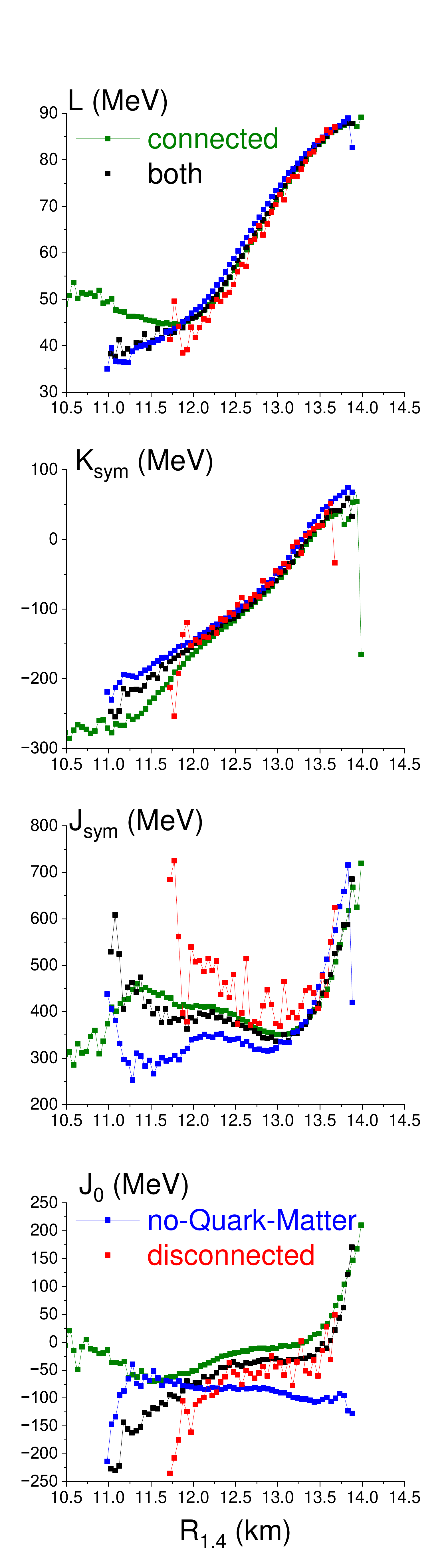}
}
\caption{
Topology-sorted inverse EOS--radius mappings of the four hadronic EOS parameters
$L$, $K_{\rm sym}$, $J_{\rm sym}$, and $J_0$. The posterior mean of each parameter is shown as a function of
the inferred canonical radius $R_{1.4}$ with $\sigma_R=0.9$ km.
}\label{HMSorted}
\end{figure}

This near universality becomes particularly clear after separating
the accepted EOSs according to their predicted posterior mass--radius topology, as shown
in Fig.~\ref{HMSorted}. The mappings of $L$ and $K_{\rm sym}$ for the
Connected, Both, and Disconnected categories largely overlap over the
radius range in which the different categories coexist. The
No-Quark-Matter category also follows essentially the same trend where
it overlaps the other classes. The small deviations at the edges of
the radius distributions (shown in the bottom panels of Fig.~\ref{QMEvery} and Fig.~\ref{QMSorted}) 
are accompanied by relatively large statistical uncertainties and should therefore not be interpreted as
strong topology dependence.

The topology independence of $L$ and $K_{\rm sym}$ has an important
physical implication. Although the four categories correspond to
qualitatively different global mass--radius sequences, they can
produce nearly the same canonical radius because the radius is
primarily controlled by the EOS at densities below or around the
onset of the high-density phase transition. Thus, knowledge of
$R_{1.4}$ can constrain the effective stiffness of the hadronic EOS
around $2\rho_0$ without uniquely determining whether the resulting
mass--radius sequence is Connected, Both, Disconnected, or contains
no stable quark-matter branch.

\subsection{High-density hadronic parameters: $J_0$ and $J_{\rm sym}$}

A qualitatively different behavior is seen for the higher-order
hadronic parameters $J_0$ and $J_{\rm sym}$. Unlike $L$ and
$K_{\rm sym}$, these parameters describe the higher-density behavior
of symmetric nuclear matter and the symmetry energy, respectively.
Within the present meta-model, they become increasingly important at
densities of several times $\rho_0$, approaching the density range in
which the hadron--quark transition can occur.

The inverse mapping of $J_{\rm sym}$ in Fig.~\ref{HMEvery} is
substantially less monotonic than those of $L$ and $K_{\rm sym}$.
Similarly, $J_0$ remains relatively weakly dependent on $R_{1.4}$ over
a broad central radius interval before changing more rapidly toward the
largest radii having extremely low probabilities to be realized. These behaviors indicate that the canonical radius
alone does not provide a simple one-to-one determination of the
high-density hadronic EOS. A given value of $R_{1.4}$ can be obtained
from substantially different combinations of high-density EOS
parameters.

This distinction is particularly important for understanding the
topology-resolved mappings in Fig.~\ref{HMSorted}. While the
$L$--$R_{1.4}$ and $K_{\rm sym}$--$R_{1.4}$ mappings remain close to
one another for the different categories, the mappings of $J_0$ and
$J_{\rm sym}$ exhibit visibly different behaviors among the
topologies. The differences are especially apparent in the lower and
intermediate radius ranges, where the Connected, Both, Disconnected,
and No-Quark-Matter EOSs occupy different regions of the high-density
parameter space.

The physical origin of this separation can be understood from the
different density scales associated with the EOS parameters. The
parameters $L$ and $K_{\rm sym}$ primarily determine the pressure
around $2\rho_0$ relevant for setting the radius of a canonical star. In contrast,
$J_0$ and $J_{\rm sym}$ influence the continuation of the hadronic EOS
to higher densities, including the stiffness of matter immediately
before the onset of deconfinement. The latter is crucial for determining
whether the appearance of quark matter destabilizes the stellar
sequence, whether stability is subsequently recovered, and whether a
connected or disconnected hybrid branch is produced.

Thus, the topology dependence of $J_0$ and $J_{\rm sym}$ does not imply
that these parameters are directly measured by $R_{1.4}$. Rather, it
shows that EOSs producing different global mass--radius topologies
require different high-density continuations even when they reproduce
essentially the same canonical radius. The topology therefore acts as
an additional discriminator of high-density EOS parameter space that
is not available from the canonical radius alone.

\subsection{Precision dependence and the Jensen interpretation}

The comparison of the $\sigma_R=0.9$ and $0.1$ km curves in
Fig.~\ref{HMEvery} provides a complementary way to interpret the
sensitivity of the inferred EOS parameters to future improvements in
radius precision. Although the numerically estimated inverse mappings obtained with the two precisions show small differences, they remain broadly consistent over the radius range of interest. These small differences reflect finite-bin, finite-sampling, and edge effects rather than a physical dependence of the underlying inverse mapping on $\sigma_R$.
Their nonlinear behavior therefore provides a natural leading-order interpretation of the precision-induced shifts through the Jensen expansion of Eq. (\ref{jensen}). To understand our results, it is important to distinguish two effects: An EOS parameter can have a steep inverse mapping and therefore be strongly correlated with $R_{1.4}$, while its mapping can nevertheless be nearly linear over
the relevant radius interval. In this case, improving the radius
precision substantially reduces the uncertainty in the inferred
parameter without necessarily producing a large systematic shift in
its posterior mean. Conversely, a parameter with a relatively weak
slope can still exhibit a noticeable precision dependence if its
inverse mapping has appreciable curvature.

The mappings of $L$ and $K_{\rm sym}$ illustrate this behavior
particularly well. Both parameters show appreciable slopes with
$R_{1.4}$, demonstrating that the canonical radius carries substantial
information about the symmetry-energy sector. Although the inverse
mappings for $\sigma_R=0.9$ and $0.1$ km remain broadly similar in the
radius range of interest, they are sufficiently nonlinear that the
finite-precision radius posterior samples the mappings differently as
$\sigma_R$ changes. Consequently, as we shall show quantitatively, improving the radius precision
produces appreciable shifts in the posterior means of both $L$ and
$K_{\rm sym}$, in addition to reducing their uncertainties. This
behavior is consistent with the Jensen expansion discussed above:
the slope of the inverse mapping characterizes the sensitivity of the
EOS parameter to the radius, while its curvature determines the
leading precision-induced shift of its posterior mean.

The behavior of $J_0$ and $J_{\rm sym}$ is different. Their mappings
contain more pronounced nonlinear structures, particularly away from
the central radius range. According to Eq.~(\ref{jensen}), such
curvature allows the finite-width radius distribution to sample
asymmetrically different parts of the inverse mapping and can
therefore shift the posterior mean. The direction of the shift is
determined by the local sign of $f_i''(\bar R)$: a locally convex mapping
produces a positive Jensen correction, whereas a locally concave
mapping produces a negative one. The smaller slopes (compared to $L$ and $K_{\rm sym}$) in the central radius region nevertheless indicate that, for the present mock observation, the radius, and thus radius-precision, dependence of these high-density hadronic parameters is not as strong as their topology dependence.

This comparison emphasizes an important distinction between
\emph{correlation with an observable} and \emph{sensitivity to
observational precision}. The slope of an inverse mapping tells us
whether the radius is informative about an EOS parameter, while its
curvature tells us whether changing the precision of the radius
measurement will shift the inferred posterior mean. The two
quantities therefore answer different questions and should not be
identified with one another.

\subsection{What the topology-resolved mappings reveal}
\label{sec:topology_mappings}
The combined information in Figs.~\ref{HMEvery} and
\ref{HMSorted} reveals a clear separation between the roles of the
hadronic EOS parameters. The symmetry-energy parameters $L$ and
$K_{\rm sym}$ are primarily \emph{radius-setting} parameters: they
control the pressure in the density range around $2\rho_0$ that largely determines
$R_{1.4}$ and consequently exhibit broadly similar inverse mappings
among the different mass--radius topologies. In contrast, the
higher-order parameters $J_0$ and $J_{\rm sym}$ govern the higher-density
continuation of the hadronic EOS and show more pronounced
topology-dependent mappings. They therefore retain information about
how the EOS approaches the hadron--quark transition and about the
global structure of the resulting mass--radius sequence.

The inverse mappings thus separate the information directly encoded in
the canonical radius from that associated with the subsequent
high-density evolution of the EOS. 
Schematically, the hierarchy of information can be represented as
\begin{eqnarray}
R_{1.4}
&\longrightarrow&
\{L,K_{\rm sym}\}
\longrightarrow
\text{allowed high-density EOS}\nonumber\\
&\longrightarrow&
\text{mass--radius topology}.\nonumber
\label{eq:informationflow}
\end{eqnarray}
This should not be interpreted as a one-to-one mapping. Rather, a
measurement of $R_{1.4}$ restricts the EOS combinations that determine
the pressure in the density range relevant for a canonical star, while
different high-density continuations can remain compatible with the
same radius. The topology-resolved mappings make this residual
degeneracy explicit by showing that EOSs associated with different
mass--radius topologies can have similar canonical radii while
occupying distinguishable regions of the higher-density parameter
space.

This distinction becomes important when the posterior PDFs and the
hadron--quark transition parameters are considered below. The mappings
identify where the topology dependence enters the EOS parameter space;
the posterior PDFs then quantify how this structure changes with the
precision of the radius measurement.

\section{Inverse Mappings of Hadron--Quark Transition Parameters}
\label{sec:QMmapping}

\begin{figure}[thb]
\centering
\vspace{-0.9cm}
 \resizebox{0.32\textwidth}{!}{
\includegraphics[width=1.0\textwidth,
 trim=0 100 0 0,
    clip
]{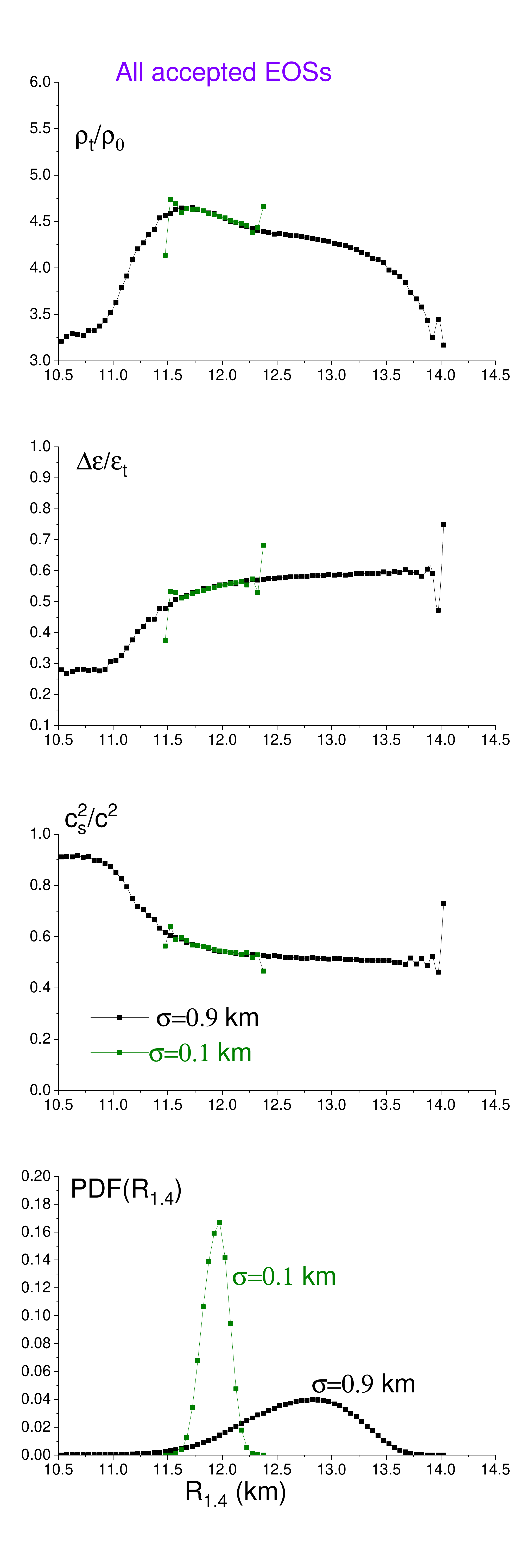}
}
\caption{
Inverse EOS--radius mappings of the hadron--quark transition density
$\rho_t/\rho_0$, normalized energy-density jump
$\Delta\varepsilon/\varepsilon_t$, and quark-matter sound speed
$c_s^2$ obtained from all accepted EOSs. The black and green curves
correspond to radius uncertainties $\sigma_R=0.9$ and $0.1$ km,
respectively. The bottom panel shows the corresponding posterior
distribution of $R_{1.4}$.
}
\label{QMEvery}
\end{figure}

We next examine how the canonical radius encodes the properties of the
hadron--quark phase transition. Fig.~\ref{QMEvery} shows the inverse
mappings of the transition density $\rho_t$, the normalized
energy-density jump $\Delta\varepsilon/\varepsilon_t$, and the
quark-matter sound speed $c_s^2$ obtained from all accepted EOSs.
In addition, the posterior PDFs of $R_{1.4}$ are also shown in the bottom panel with both 
$\sigma_R=0.9$ and $0.1$ km. In contrast to the hadronic parameters discussed above, these three
quantities play qualitatively different roles in determining the
stellar sequence. The transition density specifies \emph{when} the
new phase appears, whereas the energy-density discontinuity and the
quark-matter sound speed determine, to a large extent, \emph{how the
stellar sequence responds} to the transition.

\subsection{Transition density: the most radius-sensitive transition parameter}

With $\sigma_R=0.9$ km, the inverse mapping of the transition density $\rho_t$ displays a
pronounced and non-monotonic dependence on $R_{1.4}$. Starting from the
smallest radii represented by the accepted EOSs, the inferred
transition density increases substantially and reaches a broad maximum
around $R_{1.4}\simeq 11.5$--$12.0$ km. It then decreases gradually
toward larger radii. The corresponding mapping obtained with
$\sigma_R=0.1$ km follows the same overall behavior, although the
posterior mean is more tightly localized around the underlying
mapping.

The non-monotonic behavior is physically significant. It demonstrates
that the canonical radius does not provide a globally one-to-one
mapping onto the transition density. Rather, different combinations of
the hadronic EOS and transition parameters can produce similar stellar
radii. Nevertheless, within the radius interval most strongly favored
by the present mock data, the transition density changes appreciably
with $R_{1.4}$. This makes $\rho_t$ substantially more accessible to
radius measurements than would be expected if it were completely
degenerate with the other parameters.

The precision dependence is also particularly informative. When the
radius uncertainty is reduced from $\sigma_R=0.9$ to $0.1$ km, the
inverse mapping is sampled over a much narrower radius interval.
Consequently, the posterior uncertainty associated with $\rho_t$ is
reduced and the inferred transition density becomes more localized.
This effect is especially relevant because $\rho_t$ determines whether
the central density of a NS is sufficient for quark matter to
appear at all. A precise radius measurement can therefore provide
information about the density at which deconfinement becomes possible,
even though it does not by itself uniquely establish the composition of
the stellar core.

This interpretation is consistent with the topology-resolved
mappings shown below in the upper panel of Fig.~\ref{QMSorted}. EOSs belonging to different topological classes
occupy substantially different $\rho_t$--$R_{1.4}$ regions, indicating
that the transition density provides an important link between the
canonical radius and the subsequent high-density structure. 
We notice that for the No-Quark-Matter category, the transition parameters are
formal parameters of the underlying EOS model. In this category, either no stable quark-matter branch is realized, or the $\rho_t$ lies above the maximum density reached in canonical NSs \citep{Grundler:2025mcz}. 

\subsection{Energy-density jump: primarily a topology-defining parameter}

The inverse mapping of the normalized energy-density jump
$\Delta\varepsilon/\varepsilon_t$ behaves quite differently. For all
accepted EOSs, the mapping rises from relatively small values at the
smallest radii and then becomes approximately flat over a broad
interval extending from roughly $R_{1.4}\simeq12$ km to
$R_{1.4}\simeq13.5$ km. Over this broad interval, substantial changes in
$R_{1.4}$ correspond to only modest changes in the mean
$\Delta\varepsilon/\varepsilon_t$.
Since there are relatively few EOS with $R_{1.4} < 12$ km, this means $\Delta\varepsilon/\varepsilon_t$ does not have a strong impact on determining the canonical radius for most EOS. Thus, reducing the radius uncertainty by nearly an order of magnitude does not produce a comparably large change in the inferred energy-density jump, as discussed in more detail in the following section.

\begin{figure}[thb]
\centering
\vspace{-1.cm}
 \resizebox{0.29\textwidth}{!}{
\includegraphics[width=\textwidth,
 trim=0 20 0 0,
    clip
]{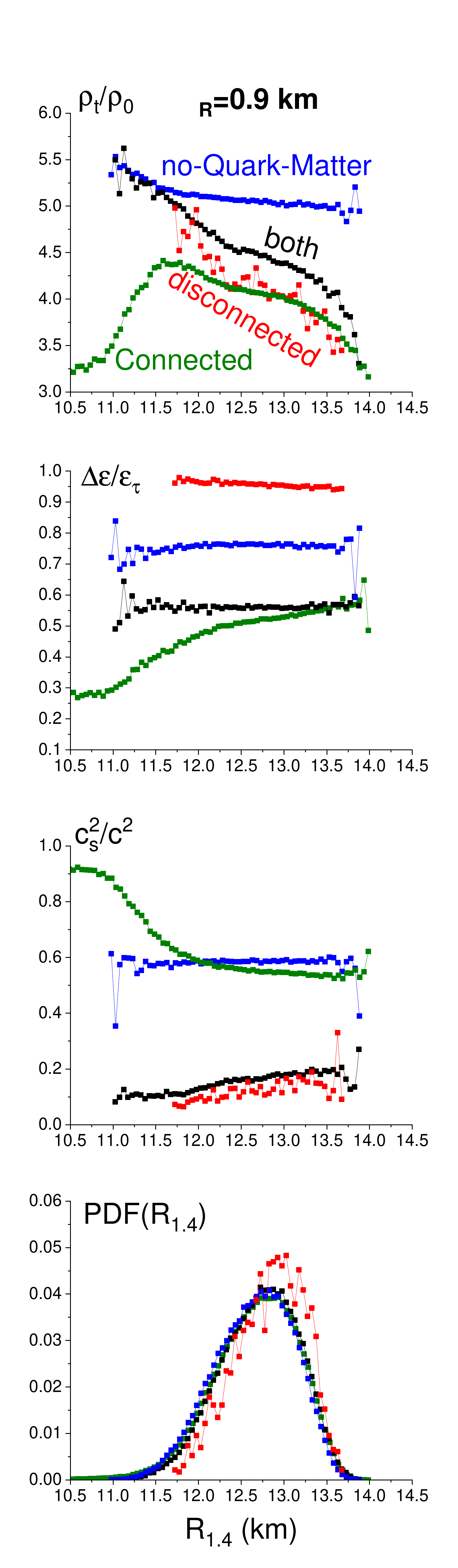}
}
\caption{Topology-sorted inverse EOS--radius mappings of the hadron--quark transition density
$\rho_t/\rho_0$, normalized energy-density jump
$\Delta\varepsilon/\varepsilon_t$, and quark-matter sound speed
$c_s^2$ obtained from all accepted EOSs with $\sigma_R=0.9$ km.
The bottom panel shows the corresponding posterior distribution of $R_{1.4}$ in different categories.
To be consistent, only the $R_{1.4}$ on the hadronic branch is used in the case of EOSs producing twin stars.
} \label{QMSorted}
\end{figure}

The topology-resolved mapping in Fig.~\ref{QMSorted} makes the reason
for this behavior much clearer. The different mass--radius topologies
occupy distinctly different ranges of
$\Delta\varepsilon/\varepsilon_t$. In particular, the Disconnected
category is concentrated near very large energy-density jumps,
whereas the Connected category extends toward substantially smaller
values. The Both category occupies an intermediate region, while the
No-Quark-Matter category also favors relatively large values.

The separation among these categories persists even though their
canonical radii overlap strongly as shown in the bottom panel. This demonstrates that
$\Delta\varepsilon/\varepsilon_t$ is primarily associated with the
\emph{global topology} of the stellar sequence rather than with the
canonical radius itself. In physical terms, a sufficiently large
energy-density discontinuity can destabilize a star when the new phase
appears through the Seidov condition \citep{seidov}, thereby creating an unstable interval and potentially a
disconnected hybrid branch. The value of the energy-density jump therefore
controls the stability response to the phase transition more directly
than it controls the radius of a canonical star.

Consequently, a high-precision measurement of $R_{1.4}$ should not be
expected to determine the energy-density jump accurately by itself. A direct
identification of this parameter will require additional information
that is sensitive to the stability structure of the stellar sequence.

\subsection{Quark-matter sound speed: a diagnostic of topology}

An even more striking behavior is obtained for the quark-matter sound
speed $c_s^2$. The inverse mapping obtained from all accepted EOSs
decreases rapidly from large values at the smallest radii and then
becomes nearly flat at $c_s^2/c^2\simeq0.5$--$0.6$ over a wide range of
$R_{1.4}$. The mappings corresponding to the two radius precisions are
again remarkably similar over the central radius range.

Thus, despite the strong physical importance of the quark-matter
sound speed for the stiffness of the high-density phase, its value is
only weakly encoded in the canonical radius. This is another example
of the composition and high-density degeneracy inherent in the TOV
mapping: different microscopic properties of matter can lead to very
similar pressure--energy-density relations over the density interval
that controls $R_{1.4}$.

The topology-resolved mapping, however, reveals a strong separation
among the different classes. The Connected category preferentially
occupies a region of relatively large $c_s^2$, while the Disconnected
category is concentrated at substantially smaller values. The Both
category lies between these extremes, and the No-Quark-Matter class
occupies a region characteristic of the absence of a stable
quark-matter branch.

This separation has a direct physical interpretation. Once quark
matter appears, the stellar sequence can remain stable only if the
high-density phase provides sufficient pressure support. A relatively
stiff quark phase, corresponding to larger $c_s^2$, can compensate for
the softening associated with the phase transition and allow a
Connected sequence to persist. In contrast, a strong softening
combined with a small quark-matter sound speed can destabilize the
star and favor the formation of a disconnected hybrid branch.

The sound speed therefore plays a role almost complementary to that of
the transition density. The latter is relatively accessible through
the radius because it determines the onset of the new phase, whereas
$c_s^2$ primarily controls what happens \emph{after} the transition.
It is consequently much more strongly associated with the topology of
the mass--radius sequence than with the value of $R_{1.4}$ itself.

\begin{table*}[htb]
\centering
\small
\caption{Posterior mean $\pm$ standard deviation of the hadronic EOS
parameters $J_0$, $J_{\rm sym}$, $K_{\rm sym}$, and $L$, together with
the three hadron--quark transition parameters, for
$\sigma_R=0.9$ and $0.1$ km.  The results are shown for the four
categories used in the present analysis. For the No-Quark-Matter category, the hadron--quark transition
parameters are formal parameters of the underlying EOS model.}
\label{tab:hadronic_qm}
\begin{tabular}{lrrrrrrr}
\hline
Category &
$J_0$ (MeV)&
$J_{\rm sym}$ (MeV)&
$K_{\rm sym}$ (MeV)&
$L$ (MeV)&
$\rho_t/\rho_0$ &
$\Delta\varepsilon/\varepsilon_t$ &
$c_s^2/c^2$
\\
\hline
\multicolumn{8}{l}{\textbf{Both}, $\sigma_R=0.9$ km}\\
& $-41.0\pm135.5$ & $369.8\pm285.3$ &
$-83.3\pm87.8$ & $63.8\pm15.3$ &
$4.48\pm0.87$ & $0.559\pm0.218$ & $0.170\pm0.227$\\
\multicolumn{8}{l}{\textbf{Both}, $\sigma_R=0.1$ km}\\
& $-82.9\pm133.1$ & $389.7\pm269.6$ &
$-158.2\pm66.2$ & $45.1\pm10.9$ &
$4.84\pm0.72$ & $0.558\pm0.225$ & $0.130\pm0.153$\\
\hline
\multicolumn{8}{l}{\textbf{Connected}, $\sigma_R=0.9$ km}\\
& $-17.7\pm149.7$ & $382.8\pm277.8$ &
$-92.4\pm93.2$ & $62.9\pm15.3$ &
$4.06\pm0.77$ & $0.511\pm0.223$ & $0.561\pm0.277$\\
\multicolumn{8}{l}{\textbf{Connected}, $\sigma_R=0.1$ km}\\
& $-54.0\pm154.4$ & $413.4\pm261.7$ &
$-172.0\pm71.7$ & $45.6\pm11.4$ &
$4.31\pm0.81$ & $0.462\pm0.210$ & $0.597\pm0.279$\\
\hline
\multicolumn{8}{l}{\textbf{No-Quark-Matter}, $\sigma_R=0.9$ km}\\
& $-85.9\pm108.3$ & $339.1\pm295.9$ &
$-80.8\pm84.0$ & $64.0\pm15.2$ &
$5.06\pm0.59$ & $0.761\pm0.170$ & $0.586\pm0.248$\\
\multicolumn{8}{l}{\textbf{No-Quark-Matter}, $\sigma_R=0.1$ km}\\
& $-79.7\pm133.6$ & $326.9\pm287.3$ &
$-148.5\pm63.8$ & $46.1\pm11.0$ &
$5.12\pm0.54$ & $0.760\pm0.167$ & $0.583\pm0.251$\\
\hline
\multicolumn{8}{l}{\textbf{Disconnected}, $\sigma_R=0.9$ km}\\
& $-55.8\pm155.3$ & $427.2\pm269.4$ &
$-63.2\pm85.3$ & $65.3\pm14.6$ &
$4.09\pm0.84$ & $0.955\pm0.034$ & $0.134\pm0.203$\\
\multicolumn{8}{l}{\textbf{Disconnected}, $\sigma_R=0.1$ km}\\
& $-127.9\pm141.2$ & $506.9\pm217.9$ &
$-154.1\pm65.3$ & $42.6\pm10.5$ &
$4.60\pm0.76$ & $0.966\pm0.028$ & $0.093\pm0.093$\\
\hline
\end{tabular}
\end{table*}

\section{POSTERIOR PDFs AND THE PRECISION DEPENDENCE OF EOS INFERENCE}

\begin{figure}[htbp]
\centering
\vspace{-0.5cm}
\includegraphics[
    width=0.4\textwidth,
    trim=0 400 0 0,
    clip
]{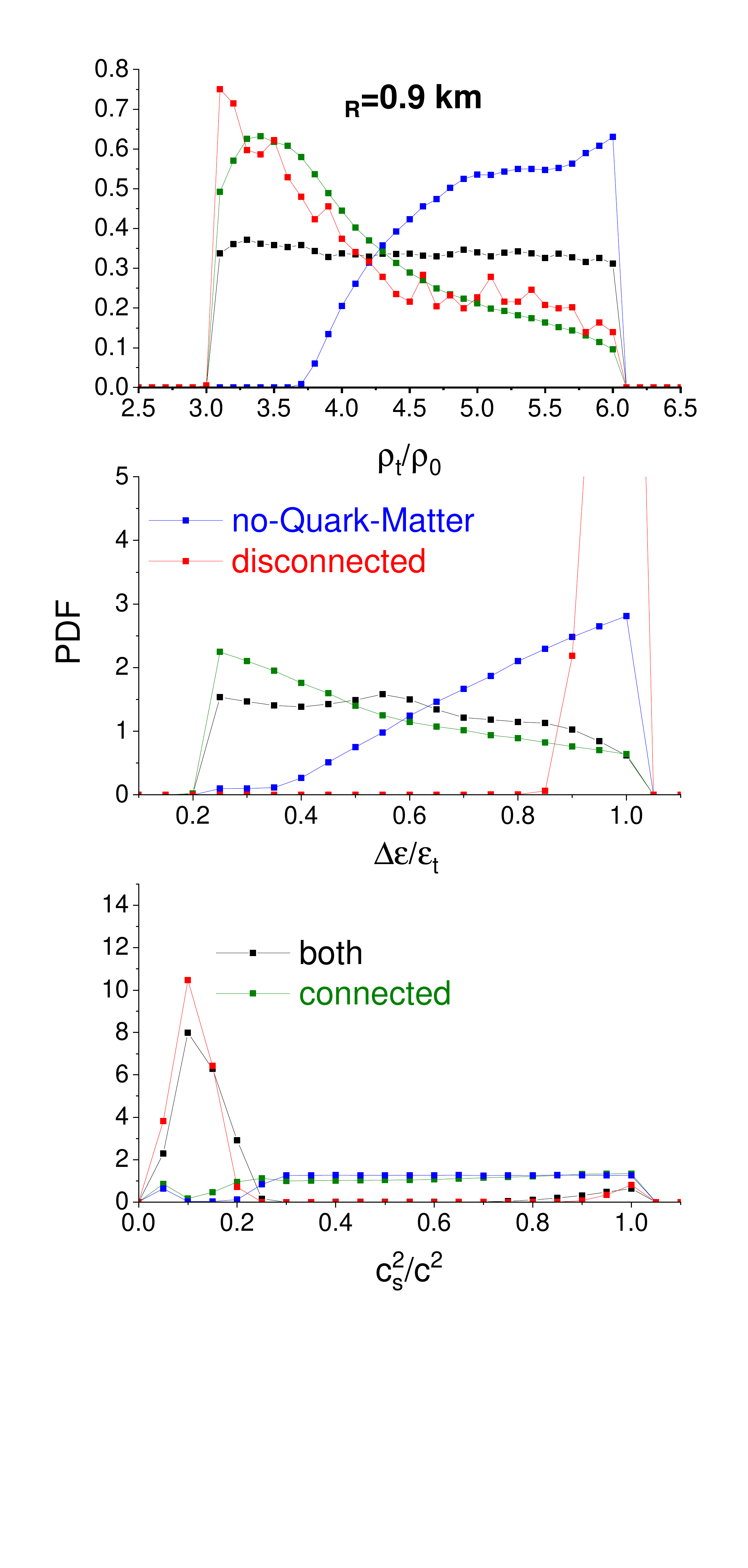}
  \setlength{\abovecaptionskip}{6pt}
\caption{Posterior PDFs of the three hadron--quark transition parameters for the four mass--radius topologies with $\sigma_R=0.9$ km.} \label{fig:QMPDF1}
\end{figure}

\begin{figure}[htbp]
\centering
\vspace{-1.cm}
 \resizebox{0.38\textwidth}{!}{
\includegraphics[width=1.0\textwidth,
 trim=0 40 0 0,
    clip
]{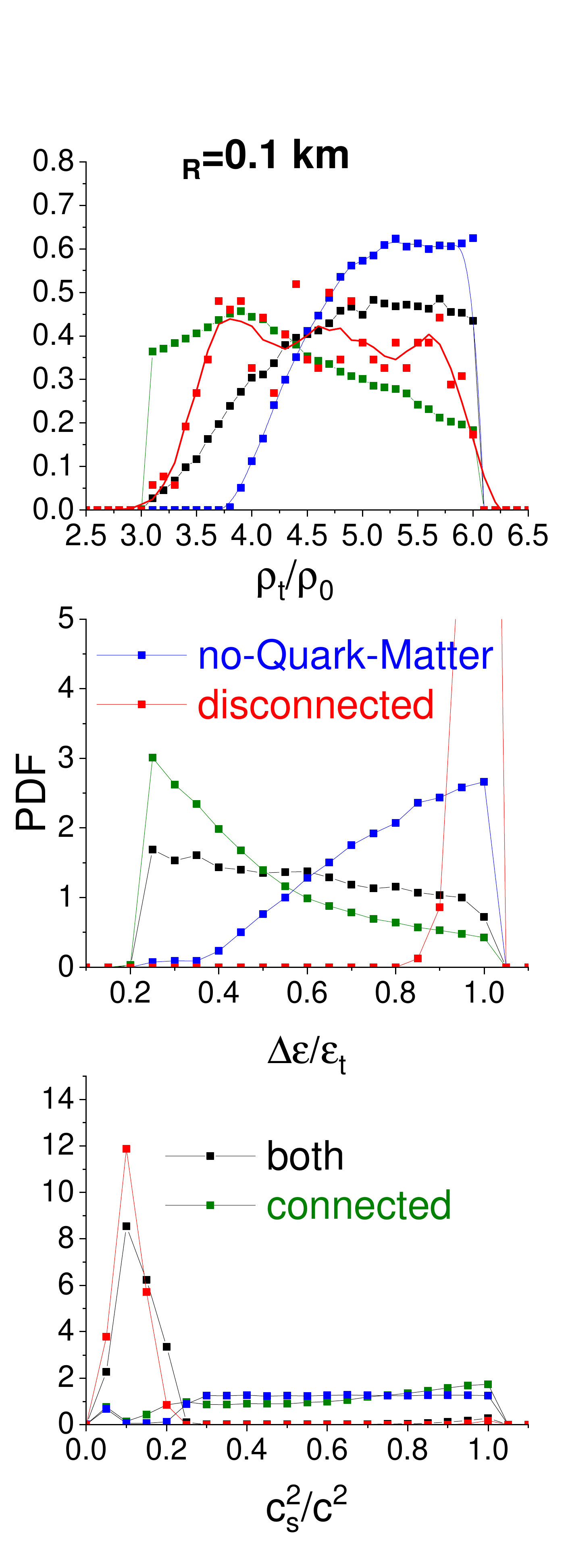}
}
\caption{Same as in Fig.~\ref{fig:QMPDF1} but with $\sigma_R=0.1$ km.} \label{fig:QMPDF2}
\end{figure}

The inverse EOS--radius mappings discussed in Sections 3 and 4 provide the physical origin of the posterior PDFs obtained from the Bayesian analyses. Here we focus on the PDFs themselves, emphasizing what changes when the uncertainty of the canonical radius measurement is reduced from $\sigma_R=0.9$ to $0.1$ km and, importantly, what does not change. Figs. \ref{fig:QMPDF1} and \ref{fig:QMPDF2} show the posterior PDFs of the three hadron--quark transition parameters for the four mass--radius topologies, while Table \ref{tab:hadronic_qm} summarizes their posterior means and standard deviations together with those of the higher-order hadronic parameters $J_0$, $J_{\rm sym}$, $K_{\rm sym}$, and $L$.
We emphasize that for the No-Quark-Matter category, these quantities should be regarded only as formal parameters of the underlying EOS model, since the
quark phase is not realized in any stable stellar configuration.

A useful point of reference is the posterior distribution
of $R_{1.4}$ itself, shown in the lower panel of Fig.~4.
Reducing $\sigma_R$ from 0.9 to 0.1 km narrows this
distribution and moves its peak toward the input value
$R_{1.4}=11.9$ km. As shown in Sec.~2.4, this behavior
arises because the Gaussian radius likelihood is multiplied
by the generally nonuniform EOS-induced density
$Q_R(R_{1.4})$. The resulting posterior therefore need
not be centered or symmetric about the measured central
value. The nonlinear TOV mapping provides the underlying
physical origin of the nonuniform distribution of model-predicted radii. In addition, the approximate
Lattimer--Prakash scaling $R_{1.4}\propto [{\rm{Pressure}(\sim 2\rho_0)}]^{1/4}$ implies
a locally nonlinear inverse relation between radius and the relevant
pressure-setting EOS sector \citep{Lattimer:2000nx}, which contributes to the
precision-dependent shifts of the inferred EOS parameters
through the Jensen expansion. Thus, the shift of the
$R_{1.4}$ posterior and the shifts of the EOS parameter
posteriors are related but should not be regarded as arising
from the same mathematical step.

\subsection{Posterior PDFs of hadronic EOS parameters}
Table \ref{tab:hadronic_qm} shows that the greatest precision-induced changes among the hadronic parameters occur for $L$ and $K_{\rm sym}$. For the Both category, for example, the posterior mean of $L$ decreases from $63.8\pm15.3$ MeV at $\sigma_R=0.9$ km to $45.1\pm10.9$ MeV at $0.1$ km, while $K_{\rm sym}$ changes from $-83.3\pm87.8$ to $-158.2\pm66.2$ MeV. Similar shifts occur in the Connected, No-Quark-Matter, and Disconnected categories. Thus, improving the radius precision substantially reduces the uncertainties of the radius-setting symmetry-energy parameters and also changes their posterior means.

The behavior of $J_0$ and $J_{\rm sym}$ is different. Their posterior means change less systematically, on average, with $\sigma_R$ than those of $L$ and $K_{\rm sym}$, while their standard deviations remain comparatively large. For the Connected category, for example, $J_0$ changes from $-17.7\pm149.7$ to $-54.0\pm154.4$ MeV and $J_{\rm sym}$ from $382.8\pm277.8$ to $413.4\pm261.7$ MeV. In the Disconnected category, the corresponding changes are larger, from $-55.8\pm155.3$ to $-127.9\pm141.2$ MeV for $J_0$ and from $427.2\pm269.4$ to $506.9\pm217.9$ MeV for $J_{\rm sym}$. These results are consistent with the inverse mappings in Figs. \ref{HMEvery} and \ref{HMSorted}: $J_0$ and $J_{\rm sym}$ probe the higher-density continuation of the hadronic EOS and are consequently less directly encoded in $R_{1.4}$ than $L$ and $K_{\rm sym}$. Their posterior distributions retain substantial width even when the radius is measured very precisely.

The topology dependence is also more apparent for $J_0$ and $J_{\rm sym}$ than for $L$ and $K_{\rm sym}$. At $\sigma_R=0.9$ km, the mean $J_{\rm sym}$ ranges from $339.1$ MeV for No-Quark-Matter to $427.2$ MeV for Disconnected EOSs, while the corresponding $J_0$ means range from $-85.9$ to $-17.7$ MeV among these categories. At $\sigma_R=0.1$ km, the separation becomes even more pronounced for $J_{\rm sym}$, reaching $326.9$--$506.9$ MeV. This behavior reinforces the conclusion from Fig. \ref{HMSorted} that higher-order hadronic parameters are not simply radius-setting quantities. Rather, they help determine the high-density continuation of the hadronic EOS and therefore retain information about the topology of the full mass--radius sequence.

\subsection{Posterior PDFs of the hadron--quark transition parameters}

The topology-resolved PDFs in Figs. \ref{fig:QMPDF1} and \ref{fig:QMPDF2} provide the clearest demonstration that the three transition parameters carry qualitatively different information. The most pronounced response to improved radius precision is found for the transition density $\rho_t$. When $\sigma_R$ decreases from $0.9$ to $0.1$ km, the PDFs of $\rho_t$ 
become generally better localized, with topology-dependent changes in both their means and widths. In particular, their means shift systematically toward larger values for the Both and Connected categories. This trend is quantified in Table \ref{tab:hadronic_qm}: the mean $\rho_t/\rho_0$ changes from $4.48\pm0.87$ to $4.84\pm0.72$ for Both and from $4.06\pm0.77$ to $4.31\pm0.81$ for Connected. The change is smaller for No-Quark-Matter, from $5.06\pm0.59$ to $5.12\pm0.54$, while the Disconnected category shifts from $4.09\pm0.84$ to $4.60\pm0.76$. Thus, among the three transition parameters, $\rho_t$ carries the clearest imprint of the precision of $R_{1.4}$.

The reason is apparent from the inverse mapping in Fig.~\ref{QMEvery}. The transition density has a substantial, although non-monotonic, dependence on $R_{1.4}$ over the region populated by the posterior. A narrower radius PDF therefore selects a narrower portion of the allowed transition-density range. Physically, $\rho_t$ determines when deconfinement first becomes possible in the stellar core and is consequently linked directly to the density scale probed by a canonical star. The posterior PDF of $\rho_t$ thus contains genuine radius information, even though the mapping is not globally one-to-one.

The normalized energy-density jump, $\Delta\epsilon/\epsilon_t$, behaves quite differently. As seen in Figs.~\ref{fig:QMPDF1} and \ref{fig:QMPDF2}, reducing $\sigma_R$ produces only modest changes within each topology. The corresponding means in Table~\ref{tab:hadronic_qm} are nearly unchanged for Both ($0.559\pm0.218$ to $0.558\pm0.225$), No-Quark-Matter ($0.761\pm0.170$ to $0.760\pm0.167$), and Disconnected ($0.955\pm0.034$ to $0.966\pm0.028$), with a somewhat larger but still moderate change for Connected ($0.511\pm0.223$ to $0.462\pm0.210$). In contrast, the PDFs remain strongly separated by topology. Disconnected configurations are concentrated near large energy-density jumps, whereas Connected configurations favor substantially smaller values. The dominant information in $\Delta\epsilon/\epsilon_t$ is therefore not the precision of the canonical radius but the stability response of the stellar sequence to the first-order transition. A sufficiently large jump can destabilize the star at the transition and thereby favor a disconnected hybrid branch.

The quark-matter sound speed $c_s^2$ shows an even clearer separation between radius precision and topology. Figures~\ref{fig:QMPDF1} and \ref{fig:QMPDF2} show only modest changes in the PDFs when $\sigma_R$ is reduced, whereas the distributions remain strongly different among the topological classes. Table~\ref{tab:hadronic_qm} illustrates this directly. For Connected EOSs the mean changes only from $0.561\pm0.277$ to $0.597\pm0.279$, while for Disconnected EOSs it decreases from $0.134\pm0.203$ to $0.093\pm0.093$. The Both category remains at relatively small values, whereas the underlying EOS models in the No-Quark-Matter category are associated with substantially larger formal values of $c_s^2/c^2$.
These differences reflect the role of $c_s^2$ in determining the stiffness of the post-transition quark phase: a sufficiently stiff quark phase can compensate for the softening at deconfinement and maintain a connected hybrid branch, while a softer phase favors destabilization and a disconnected branch.

\subsection{Precision dependence versus topology dependence}

Figures~\ref{fig:QMPDF1} and \ref{fig:QMPDF2} and Table~\ref{tab:hadronic_qm} highlight an important distinction between
the effects of radius precision and mass--radius topology.  The transition
density $\rho_t$ retains appreciable sensitivity to $R_{1.4}$ and therefore
responds measurably to improved radius precision.  In contrast, the
normalized energy-density jump $\Delta\epsilon/\epsilon_t$ and the
quark-matter sound speed $c_s^2/c^2$ change only modestly with $\sigma_R$
within a given topology, while their posterior distributions remain
strongly separated among the different topological classes.  The
energy-density jump characterizes the strength of the softening at the
first-order transition, whereas $c_s^2/c^2$ controls the stiffness of the
post-transition phase; both therefore leave a stronger imprint on the
stability and topology of the full stellar sequence than on $R_{1.4}$
alone.

The combined results show that the gain from improving radius precision
is intrinsically parameter dependent.  High-precision measurements of
$R_{1.4}$ most directly improve the inference of the radius-setting
parameters $L$ and $K_{\rm sym}$ and provide useful additional information
on the transition density $\rho_t$.  By contrast, the strength of the
transition and the stiffness of the quark phase remain more strongly tied
to the global mass--radius topology.  Thus, precise radius measurements
and observations capable of identifying the topology provide
complementary information: the former constrains the EOS leading toward
the transition, whereas the latter helps determine the physical nature
and consequences of the transition.

\section{SUMMARY AND CONCLUSIONS}
\label{sec:summary}

We have investigated how future high-precision measurements of the
canonical NS radius can constrain the dense-matter EOS and
the properties of a possible first-order hadron--quark transition.
Using a Bayesian meta-model with nine EOS parameters and mock
measurements $R_{1.4}=11.9\pm\sigma_R$ km for
$\sigma_R=0.9$ and $0.1$ km, we introduced inverse EOS--radius mappings by characterizing the conditional mean of each EOS parameter as a function of the model-predicted canonical radius $R_{1.4}$.
The slope of this mapping characterizes the sensitivity of an EOS parameter to the radius, while its curvature determines the leading precision dependence of its posterior mean through the Jensen expansion.

The mappings reveal a hierarchy in the information encoded by
$R_{1.4}$. The symmetry-energy parameters $L$ and $K_{\rm sym}$ are
strongly correlated with the canonical radius, and their posterior
means shift appreciably as the radius precision improves. In contrast,
$J_0$ and $J_{\rm sym}$ show weaker precision dependence but more
pronounced topology dependence, reflecting the freedom of the
high-density continuation of the hadronic EOS. Thus, EOSs with similar
$R_{1.4}$ can occupy substantially different regions of high-density
parameter space.

The same distinction is particularly clear for the hadron--quark
transition parameters. The transition density $\rho_t$ 
retains substantial radius sensitivity and is generally better constrained as the radius precision improves.
In contrast, the normalized energy-density jump
$\Delta\epsilon/\epsilon_t$ and the quark-matter sound speed
$c_s^2/c^2$ show relatively weak precision dependence but strong
separation among the four mass--radius topologies. They are therefore
more directly related to the stability and global topology of the
stellar sequence than to the canonical radius itself.

Importantly, the topology-resolved posterior distributions of
$R_{1.4}$ overlap strongly. A precise canonical radius therefore cannot
by itself identify whether the sequence is Connected, Disconnected,
Both, or No-Quark-Matter, nor can it uniquely determine the
high-density transition properties. Instead, $R_{1.4}$ primarily
constrains the pressure-setting part of the EOS and, through its
correlations with the high-density EOS, provides useful information on
the transition density. Additional observations sensitive to the
global high-density stellar structure are required to distinguish the
topologies and constrain the strength and post-transition stiffness of
a possible hadron--quark transition.

Overall, the inverse mappings provide a physical interpretation of the
posterior PDFs and establish that the scientific return of improving
radius precision is intrinsically parameter dependent. High-precision
radius measurements are especially valuable for the EOS components
that control the canonical NS radius and for the onset density of
deconfinement, while complementary probes of the global mass--radius
structure are needed to determine the nature and consequences of a
possible hadron--quark transition.
We emphasize that these conclusions are conditional on the
adopted $3\leq\rho_t/\rho_0\leq6$ transition-density prior,
the use of the first (hadronic) branch for $R_{1.4}$, the
conventional rapid-conversion stability criterion, and the
assumption of a single hadron--quark transition.
\\

\noindent{\bf Acknowledgements.} This work is supported in part by the U.S. Department of Energy, Office of Science, under Award Number DE-SC0013702, and the NASA-Texas Space Grant Consortium.

\bibliographystyle{plainnat}
\bibliographystyle{aasjournal}
\bibliography{references}

@article{Pereira:2017rmp,
    author = "Pereira, Jonas P. and Flores, C{\'e}sar V. and Lugones, Germ{\'a}n",
    title = "{Phase transition effects on the dynamical stability of hybrid neutron stars}",
    eprint = "1706.09371",
    archivePrefix = "arXiv",
    primaryClass = "gr-qc",
    doi = "10.3847/1538-4357/aabfbf",
    journal = "Astrophys. J.",
    volume = "860",
    number = "1",
    pages = "12",
    year = "2018"
}

@article{Rau:2022ofy,
    author = "Rau, Peter B. and Sedrakian, Armen",
    title = "{Two first-order phase transitions in hybrid compact stars: Higher-order multiplet stars, reaction modes, and intermediate conversion speeds}",
    eprint = "2212.09828",
    archivePrefix = "arXiv",
    primaryClass = "astro-ph.HE",
    reportNumber = "INT-PUB-22-056",
    doi = "10.1103/PhysRevD.107.103042",
    journal = "Phys. Rev. D",
    volume = "107",
    number = "10",
    pages = "103042",
    year = "2023"
}

@article{Alford:2015dpa,
    author = "Alford, Mark G. and Burgio, G. F. and Han, Sophia and Taranto, Gabriele and Zappal{\`a}, Dario",
    title = "{Constraining and applying a generic high-density equation of state}",
    eprint = "1501.07902",
    archivePrefix = "arXiv",
    primaryClass = "nucl-th",
    doi = "10.1103/PhysRevD.92.083002",
    journal = "Phys. Rev. D",
    volume = "92",
    number = "8",
    pages = "083002",
    year = "2015"
}

@article{Alford:2015gna,
    author = "Alford, Mark G. and Han, Sophia",
    title = "{Characteristics of hybrid compact stars with a sharp hadron-quark interface}",
    eprint = "1508.01261",
    archivePrefix = "arXiv",
    primaryClass = "nucl-th",
    doi = "10.1140/epja/i2016-16062-9",
    journal = "Eur. Phys. J. A",
    volume = "52",
    number = "3",
    pages = "62",
    year = "2016"
}

@article{Alford:2017qgh,
    author = "Alford, Mark G. and Sedrakian, Armen",
    title = "{Compact stars with sequential QCD phase transitions}",
    eprint = "1706.01592",
    archivePrefix = "arXiv",
    primaryClass = "astro-ph.HE",
    doi = "10.1103/PhysRevLett.119.161104",
    journal = "Phys. Rev. Lett.",
    volume = "119",
    number = "16",
    pages = "161104",
    year = "2017"
}

@article{Li:2026hnz,
    author = "Li, Bao-An and Grundler, Xavier",
    title = "{Quantifying the information gain from future high-precision radius measurements for identifying twin neutron stars}",
    eprint = "2607.18124",
    archivePrefix = "arXiv",
    primaryClass = "astro-ph.HE",
    doi = "10.1016/j.physletb.2026.140863",
    journal = "Phys. Lett. B",
    volume = "880",
    pages = "140863",
    year = "2026"
}

@article{Haque:2026dre,
    author = "Haque, Shamim and Shinde, Atharva and Saha, Asim Kumar and Malik, Tuhin and Mallick, Ritam",
    title = "{Investigating Twin Star Equation of States in Light of Recent Astrophysical Observations}",
    eprint = "2601.16674",
    archivePrefix = "arXiv",
    primaryClass = "astro-ph.HE",
    month = "1",
    year = "2026"
}

@article{Haque26,
  title = {Dynamical response of twin stars to perturbations},
  author = {Haque, Shamim and Rezzolla, Luciano and Mallick, Ritam},
  journal = {Phys. Rev. D},
  volume = {113},
  issue = {10},
  pages = {103044},
  numpages = {12},
  year = {2026},
  month = {May},
  publisher = {American Physical Society},
  doi = {10.1103/99c1-2p2y},
  url = {https://link.aps.org/doi/10.1103/99c1-2p2y}
}

@article{Tsa23,
  title = {Twin stars as probes of the nuclear equation of state: Effects of rotation through the PSR J0952-0607 pulsar and constraints via the tidal deformability from the GW170817 event},
  author = {Tsaloukidis, Lazaros and Koliogiannis, P. S. and Kanakis-Pegios, A. and Moustakidis, Ch. C.},
  journal = {Phys. Rev. D},
  volume = {107},
  issue = {2},
  pages = {023012},
  numpages = {18},
  year = {2023},
  month = {Jan},
  publisher = {American Physical Society},
  doi = {10.1103/PhysRevD.107.023012},
  url = {https://link.aps.org/doi/10.1103/PhysRevD.107.023012}
}

@article{Christian_2018,
    author = {Christian, Jan-Erik and Zacchi, Andreas and Schaffner-Bielich, J{\"u}rgen},
    title = "{Classifications of Twin Star Solutions for a Constant Speed of Sound Parameterized Equation of State}",
    eprint = "1707.07524",
    archivePrefix = "arXiv",
    primaryClass = "astro-ph.HE",
    doi = "10.1140/epja/i2018-12472-y",
    journal = "Eur. Phys. J. A",
    volume = "54",
    number = "2",
    pages = "28",
    year = "2018"
}

@ARTICLE{Pal_2025,
       author = {{Pal}, Suman and {Podder}, Soumen and {Chaudhuri}, Gargi},
        title = {Is the Central Compact Object in {HESS J1731-347} a Hybrid Star with a Quark Core? {An} Analysis with the Constant Speed of Sound Parameterization},
      journal = {Astrophysical Journal},
         year = 2025,
        month = apr,
       volume = {983},
       number = {1},
          eid = {24},
        pages = {24},
          doi = {10.3847/1538-4357/adbc6b},
archivePrefix = {arXiv},
       eprint = {2504.02945},
 primaryClass = {nucl-th},
       adsurl = {https://ui.adsabs.harvard.edu/abs/2025ApJ...983...24P}
}

@article{Laskos_Patkos_2025,
  title = {XTE J1814-338: A potential hybrid star candidate},
  author = {Laskos-Patkos, P. and Moustakidis, Ch. C.},
  journal = {Phys. Rev. D},
  volume = {111},
  issue = {6},
  pages = {063058},
  numpages = {8},
  year = {2025},
  month = {Mar},
  publisher = {American Physical Society},
  doi = {10.1103/PhysRevD.111.063058},
  url = {https://link.aps.org/doi/10.1103/PhysRevD.111.063058}
}

@article{Christian_2025,
    author = "Christian, Jan-Erik and Rather, Ishfaq Ahmad and Gholami, Hosein and Hofmann, Marco",
    title = "{Comprehensive analysis of constructing hybrid stars with a renormalization group-consistent Nambu-Jona-Lasino model}",
    eprint = "2503.13626",
    archivePrefix = "arXiv",
    primaryClass = "astro-ph.HE",
    doi = "10.1051/0004-6361/202555009",
    journal = "Astron. Astrophys.",
    volume = "701",
    pages = "A145",
    year = "2025"
}

@ARTICLE{huang2025,
       author = {{Huang}, Chun and {Sourav}, Shashwat},
        title = {Constraining First-order Phase Transition inside Neutron Stars with Application of Bayesian Techniques on {PSR J0437{\textendash}4715 NICER} Data},
      journal = {Astrophysical Journal},
         year = 2025,
        month = apr,
       volume = {983},
       number = {1},
          eid = {17},
        pages = {17},
          doi = {10.3847/1538-4357/adbb67},
archivePrefix = {arXiv},
       eprint = {2502.11976},
 primaryClass = {astro-ph.HE},
       adsurl = {https://ui.adsabs.harvard.edu/abs/2025ApJ...983...17H}
}

@ARTICLE{Gorda_2023,
       author = {{Gorda}, T. and {Hebeler}, K. and {Kurkela}, A. and {Schwenk}, A. and {Vuorinen}, A.},
        title = {Constraints on Strong Phase Transitions in Neutron Stars},
      journal = {Astrophysical Journal},
         year = 2023,
        month = oct,
       volume = {955},
       number = {2},
          eid = {100},
        pages = {100},
          doi = {10.3847/1538-4357/aceefb},
archivePrefix = {arXiv},
       eprint = {2212.10576},
 primaryClass = {astro-ph.HE},
       adsurl = {https://ui.adsabs.harvard.edu/abs/2023ApJ...955..100G}
}

@article{Albino:2024ymc,
    author = "Albino, Milena and Malik, Tuhin and Ferreira, M\'arcio and Provid\^encia, Constan\c{c}a",
    title = "{Hybrid star properties with the NJL and mean field approximation of QCD models: A Bayesian approach}",
    eprint = "2406.15337",
    archivePrefix = "arXiv",
    primaryClass = "nucl-th",
    doi = "10.1103/PhysRevD.110.083037",
    journal = "Phys. Rev. D",
    volume = "110",
    number = "8",
    pages = "083037",
    year = "2024"
}

@article{Li_2024,
   title={Hybrid Star Models in the Light of New Multimessenger Data},
   volume={967},
   ISSN={1538-4357},
   url={http://dx.doi.org/10.3847/1538-4357/ad4295},
   DOI={10.3847/1538-4357/ad4295},
   number={2},
   journal={The Astrophysical Journal},
   publisher={American Astronomical Society},
   author={Li, Jia Jie and Sedrakian, Armen and Alford, Mark},
   year={2024},
   month=may, pages={116} }

@article{Alvarez_Castillo_2016,
    author = "Alvarez-Castillo, D. and Ayriyan, A. and Benic, S. and Blaschke, D. and Grigorian, H. and Typel, S.",
    title = "{New class of hybrid EoS and Bayesian M-R data analysis}",
    eprint = "1603.03457",
    archivePrefix = "arXiv",
    primaryClass = "nucl-th",
    doi = "10.1140/epja/i2016-16069-2",
    journal = "Eur. Phys. J. A",
    volume = "52",
    number = "3",
    pages = "69",
    year = "2016"
}

@article{gerlach_1968,
  title = {Equation of State at Supranuclear Densities and the Existence of a Third Family of Superdense Stars},
  author = {Gerlach, Ulrich H.},
  journal = {Phys. Rev.},
  volume = {172},
  issue = {5},
  pages = {1325--1330},
  numpages = {0},
  year = {1968},
  month = {Aug},
  publisher = {American Physical Society},
  doi = {10.1103/PhysRev.172.1325},
  url = {https://link.aps.org/doi/10.1103/PhysRev.172.1325}
}

@article{Kampfer:1981yr,
    author = "Kampfer, Burkhard",
    title = "{On the Possibility of Stable Quark and Pion Condensed Stars}",
    doi = "10.1088/0305-4470/14/11/009",
    journal = "J. Phys. A",
    volume = "14",
    pages = "L471--L475",
    year = "1981"
}

@article{Glendenning:1998ag,
    author = "Glendenning, Norman K. and Kettner, Christiane",
    title = "{Nonidentical neutron star twins}",
    eprint = "astro-ph/9807155",
    archivePrefix = "arXiv",
    reportNumber = "LBL-42080, LBNL-42080",
    journal = "Astron. Astrophys.",
    volume = "353",
    pages = "L9",
    year = "2000"
}

@article{Schertler_2000,
   title={Quark phases in neutron stars and a third family of compact stars as signature for phase transitions},
   volume={677},
   ISSN={0375-9474},
   url={http://dx.doi.org/10.1016/S0375-9474(00)00305-5},
   DOI={10.1016/s0375-9474(00)00305-5},
   number={1–4},
   journal={Nuclear Physics A},
   publisher={Elsevier BV},
   author={Schertler, K. and Greiner, C. and Schaffner-Bielich, J. and Thoma, M.H.},
   year={2000},
   month=sep,
   pages={463–490}
}

@article{Chanlaridis:2024rov,
    author = "Chanlaridis, S. and Ohse, D. and Alvarez-Castillo, D. E. and Antoniadis, J. and Blaschke, D. and Danchev, V. and Langer, N. and Misra, D.",
    title = "{Formation of twin compact stars in low-mass X-ray binaries - Implications for eccentric and isolated millisecond pulsar populations}",
    eprint = "2409.04755",
    archivePrefix = "arXiv",
    primaryClass = "astro-ph.HE",
    doi = "10.1051/0004-6361/202452259",
    journal = "Astron. Astrophys.",
    volume = "695",
    pages = "A16",
    year = "2025"
}

@article{Veselsky:2024bnf,
    author = "Veselsky, M. and Petousis, V. and Koliogiannis, P. S. and Moustakidis, Ch. C. and Leja, J.",
    title = "{Simultaneous explanation of XTE J1814-338 and HESS J1731-347 objects using $K^-$ and $\bar{K}^0$ condensates}",
    eprint = "2412.01426",
    archivePrefix = "arXiv",
    primaryClass = "nucl-th",
    doi = "10.1103/PhysRevD.111.L061308",
    journal = "Phys. Rev. D",
    volume = "111",
    number = "6",
    pages = "L061308",
    year = "2025"
}

@article{Carlomagno:2023nrc,
    author = "Carlomagno, J. P. and Contrera, G. A. and Grunfeld, A. G. and Blaschke, D.",
    title = "{Thermal twin stars within a hybrid equation of state based on a nonlocal chiral quark model compatible with modern astrophysical observations}",
    eprint = "2312.01975",
    archivePrefix = "arXiv",
    primaryClass = "nucl-th",
    doi = "10.1103/PhysRevD.109.043050",
    journal = "Phys. Rev. D",
    volume = "109",
    number = "4",
    pages = "043050",
    year = "2024"
}

@article{Jimenez:2024hib,
    author = "Jim\'enez, Jos\'e C. and Lazzari, Lucas and Gon\c{c}alves, Victor P.",
    title = "{How the QCD trace anomaly behaves at the core of twin stars?}",
    eprint = "2408.11614",
    archivePrefix = "arXiv",
    primaryClass = "hep-ph",
    doi = "10.1103/PhysRevD.110.114014",
    journal = "Phys. Rev. D",
    volume = "110",
    number = "11",
    pages = "114014",
    year = "2024"
}

@article{Antoniadis:2013pzd,
    author = "Antoniadis, John and others",
    title = "{A Massive Pulsar in a Compact Relativistic Binary}",
    eprint = "1304.6875",
    archivePrefix = "arXiv",
    primaryClass = "astro-ph.HE",
    doi = "10.1126/science.1233232",
    journal = "Science",
    volume = "340",
    pages = "6131",
    year = "2013"
}

@article{Benic:2014jia,
    author = "Benic, Sanjin and Blaschke, David and Alvarez-Castillo, David E. and Fischer, Tobias and Typel, Stefan",
    title = "{A new quark-hadron hybrid equation of state for astrophysics - I. High-mass twin compact stars}",
    eprint = "1411.2856",
    archivePrefix = "arXiv",
    primaryClass = "astro-ph.HE",
    reportNumber = "ZTF-EP-14-13",
    doi = "10.1051/0004-6361/201425318",
    journal = "Astron. Astrophys.",
    volume = "577",
    pages = "A40",
    year = "2015"
}

@article{Kaltenborn:2017hus,
    author = "Kaltenborn, Mark Alexander Randolph and Bastian, Niels-Uwe Friedrich and Blaschke, David Bernhard",
    title = "{Quark-nuclear hybrid star equation of state with excluded volume effects}",
    eprint = "1701.04400",
    archivePrefix = "arXiv",
    primaryClass = "astro-ph.HE",
    doi = "10.1103/PhysRevD.96.056024",
    journal = "Phys. Rev. D",
    volume = "96",
    number = "5",
    pages = "056024",
    year = "2017"
}

@article{Alvarez-Castillo:2018pve,
    author = "Alvarez-Castillo, D. E. and Blaschke, D. B. and Grunfeld, A. G. and Pagura, V. P.",
    title = "{Third family of compact stars within a nonlocal chiral quark model equation of state}",
    eprint = "1805.04105",
    archivePrefix = "arXiv",
    primaryClass = "hep-ph",
    doi = "10.1103/PhysRevD.99.063010",
    journal = "Phys. Rev. D",
    volume = "99",
    number = "6",
    pages = "063010",
    year = "2019"
}

@article{Oertel:2016bki,
    author = {Oertel, M. and Hempel, M. and Kl{\"a}hn, T. and Typel, S.},
    title = "{Equations of state for supernovae and compact stars}",
    eprint = "1610.03361",
    archivePrefix = "arXiv",
    primaryClass = "astro-ph.HE",
    doi = "10.1103/RevModPhys.89.015007",
    journal = "Rev. Mod. Phys.",
    volume = "89",
    number = "1",
    pages = "015007",
    year = "2017"
}

@article{Baiotti:2019sew,
    author = "Baiotti, Luca",
    title = "{Gravitational waves from neutron star mergers and their relation to the nuclear equation of state}",
    eprint = "1907.08534",
    archivePrefix = "arXiv",
    primaryClass = "astro-ph.HE",
    doi = "10.1016/j.ppnp.2019.103714",
    journal = "Prog. Part. Nucl. Phys.",
    volume = "109",
    pages = "103714",
    year = "2019"
}

@article{Lattimer:2021emm,
    author = "Lattimer, J. M.",
    title = "{Neutron Stars and the Nuclear Matter Equation of State}",
    doi = "10.1146/annurev-nucl-102419-124827",
    journal = "Ann. Rev. Nucl. Part. Sci.",
    volume = "71",
    pages = "433--464",
    year = "2021"
}

@article{Chatziioannou:2024jsr,
    author = "Chatziioannou, Katerina and Cromartie, H. Thankful and Gandolfi, Stefano and Tews, Ingo and Radice, David and Steiner, Andrew W. and Watts, Anna L.",
    title = "{Neutron stars and the dense matter equation of state}",
    eprint = "2407.11153",
    archivePrefix = "arXiv",
    primaryClass = "nucl-th",
    reportNumber = "LA-UR-23-22545",
    doi = "10.1103/ymsq-cfcw",
    journal = "Rev. Mod. Phys.",
    volume = "97",
    number = "4",
    pages = "045007",
    year = "2025"
}

@article{Walker:2024loo,
    author = "Walker, Kris and Smith, Rory and Thrane, Eric and Reardon, Daniel J.",
    title = "{Precision constraints on the neutron star equation of state with third-generation gravitational-wave observatories}",
    eprint = "2401.02604",
    archivePrefix = "arXiv",
    primaryClass = "astro-ph.HE",
    reportNumber = "LIGO-P2300446",
    doi = "10.1103/PhysRevD.110.043013",
    journal = "Phys. Rev. D",
    volume = "110",
    number = "4",
    pages = "043013",
    year = "2024"
}

@article{Cruise,
    author = "Cruise, Mike and others",
    title = "{The NewAthena mission concept in the context of the next decade of X-ray astronomy}",
    eprint = "2501.03100",
    archivePrefix = "arXiv",
    primaryClass = "astro-ph.IM",
    doi = "10.1038/s41550-024-02416-3",
    journal = "Nature Astron.",
    volume = "9",
    number = "1",
    pages = "36--44",
    year = "2025"
}

@article{Jen,
    author = "J.~L.~W.~V.~Jensen",
    title = "Sur les fonctions convexes et les inégalités entre les valeurs moyennes",
    journal ="Acta Math.",
    doi = "10.1007/BF02418571",
    volume = "30",
    number = "11",
    pages = "175-193",
    year =1906, 
}

@article{Ber,
    author ="D.~Bernoulli" ,
    title = "Exposition of a New Theory on the Measurement of Risk",
    url="https://www.jstor.org/stable/1909829",
    journal = "Econometrica ",
    volume="22",
    pages="23",
    year = 1954
}

@book{infobook,
    author ="T. M. Cover and J. A. Thomas",
    title ="Elements of Information Theory",
    publisher ="Wiley",
    doi="10.1002/047174882X",
    url="https://doi.org/10.1002/047174882X",
    year =2006 
}

@book{Ebook,
  author    = {Mas-Colell, Andreu and Whinston, Michael D. and Green, Jerry R.},
  title     = {Microeconomic Theory},
  publisher = {Oxford University Press},
  year      = {1995},
  url="https://global.oup.com/academic/product/microeconomic-theory-9780195073409",
  isbn      = {978-0195073409}
}

@article{STAR:2025owm,
    author = "Aboona, B. E. and others",
    collaboration = "STAR",
    title = "{Onset of Constituent Quark Number Scaling in Heavy-Ion Collisions at RHIC}",
    eprint = "2504.02531",
    archivePrefix = "arXiv",
    primaryClass = "nucl-ex",
    doi = "10.1103/2qhx-cp79",
    journal = "Phys. Rev. Lett.",
    volume = "135",
    number = "7",
    pages = "072301",
    year = "2025"
}

@article{AngLi25,
    author = "Li, Ang and others",
    title = "{Dense matter in neutron stars with eXTP}",
    eprint = "2506.08104",
    archivePrefix = "arXiv",
    primaryClass = "astro-ph.HE",
    doi = "10.1007/s11433-025-2761-4",
    journal = "Sci. China Phys. Mech. Astron.",
    volume = "68",
    number = "11",
    pages = "119503",
    year = "2025"
}

@article{zdunik2013,
    author = {{Zdunik, J. L.} and {Haensel, P.}},
    title = {Maximum mass of neutron stars and strange neutron-star cores},
    DOI= "10.1051/0004-6361/201220697",
    url= "https://doi.org/10.1051/0004-6361/201220697",
    journal = {Astronomy \& Astrophysics},
    year = 2013,
    volume = 551,
    pages = "A61",
    month = "",
}

@article{STAR2,
    author = "Abdallah, M. S. and others",
    collaboration = "STAR",
    title = "{Measurements of Proton High Order Cumulants in $\sqrt{s_{_{\mathrm{NN}}}}$ = 3 GeV Au+Au Collisions and Implications for the QCD Critical Point}",
    eprint = "2112.00240",
    archivePrefix = "arXiv",
    primaryClass = "nucl-ex",
    doi = "10.1103/PhysRevLett.128.202303",
    journal = "Phys. Rev. Lett.",
    volume = "128",
    number = "20",
    pages = "202303",
    year = "2022"
}

@article{Yong22,
    author = "Yong, Gao-Chan and Li, Bao-An and Xiao, Zhi-Gang and Lin, Zi-Wei",
    title = "{Probing the high-density nuclear symmetry energy with the \ensuremath{\Xi}\ensuremath{-}/\ensuremath{\Xi}0 ratio in heavy-ion collisions at $\sqrt{s_{NN}}$\ensuremath{\approx}3 GeV}",
    eprint = "2206.10766",
    archivePrefix = "arXiv",
    primaryClass = "nucl-th",
    doi = "10.1103/PhysRevC.106.024902",
    journal = "Phys. Rev. C",
    volume = "106",
    number = "2",
    pages = "024902",
    year = "2022"
}

@article{ART,
    author = "Li, Bao-An and Ko, Che Ming",
    title = "{Formation of superdense hadronic matter in high-energy heavy ion collisions}",
    eprint = "nucl-th/9505016",
    archivePrefix = "arXiv",
    doi = "10.1103/PhysRevC.52.2037",
    journal = "Phys. Rev. C",
    volume = "52",
    pages = "2037--2063",
    year = "1995"
}

@article{LCK,
    author = "Li, Bao-An and Chen, Lie-Wen and Ko, Che Ming",
    title = "{Recent Progress and New Challenges in Isospin Physics with Heavy-Ion Reactions}",
    eprint = "0804.3580",
    archivePrefix = "arXiv",
    primaryClass = "nucl-th",
    doi = "10.1016/j.physrep.2008.04.005",
    journal = "Phys. Rept.",
    volume = "464",
    pages = "113--281",
    year = "2008"
}

@article{Zhang:2023wqj,
    author = "Zhang, Nai-Bo and Li, Bao-An",
    title = "{Properties of first-order hadron-quark phase transition from inverting neutron star observables}",
    eprint = "2304.07381",
    archivePrefix = "arXiv",
    primaryClass = "nucl-th",
    doi = "10.1103/PhysRevC.108.025803",
    journal = "Phys. Rev. C",
    volume = "108",
    number = "2",
    pages = "025803",
    year = "2023"
}

@article{Xie:2024mxu,
    author = "Xie, Wen-Jie and Li, Bao-An and Zhang, Nai-Bo",
    title = "{Impact of the newly revised gravitational redshift of x-ray burster GS 1826-24 on the equation of state of supradense neutron-rich matter}",
    eprint = "2404.01989",
    archivePrefix = "arXiv",
    primaryClass = "astro-ph.HE",
    doi = "10.1103/PhysRevD.110.043025",
    journal = "Phys. Rev. D",
    volume = "110",
    number = "4",
    pages = "043025",
    year = "2024"
}

@article{Li:2024imk,
    author = "Li, Bao-An and Grundler, Xavier and Xie, Wen-Jie and Zhang, Nai-Bo",
    title = "{Bayesian inference of fine features of the nuclear equation of state from future neutron star radius measurements to 0.1~km accuracy}",
    eprint = "2407.07823",
    archivePrefix = "arXiv",
    primaryClass = "astro-ph.HE",
    doi = "10.1103/PhysRevD.110.103040",
    journal = "Phys. Rev. D",
    volume = "110",
    number = "10",
    pages = "103040",
    year = "2024"
}

@article{Alford:2013aca,
    author = "Alford, Mark G. and Han, Sophia and Prakash, Madappa",
    title = "{Generic conditions for stable hybrid stars}",
    eprint = "1302.4732",
    archivePrefix = "arXiv",
    primaryClass = "astro-ph.SR",
    doi = "10.1103/PhysRevD.88.083013",
    journal = "Phys. Rev. D",
    volume = "88",
    number = "8",
    pages = "083013",
    year = "2013"
}

@article{Zhang:2020zsc,
    author = "Zhang, Nai-Bo and Li, Bao-An",
    title = "{GW190814's Secondary Component with Mass 2.50\textendash{}2.67 M $_{\odot}$ as a Superfast Pulsar}",
    eprint = "2007.02513",
    archivePrefix = "arXiv",
    primaryClass = "astro-ph.HE",
    doi = "10.3847/1538-4357/abb470",
    journal = "Astrophys. J.",
    volume = "902",
    number = "1",
    pages = "38",
    year = "2020"
}

@article{Zhang:2018bwq,
    author = "Zhang, Nai-Bo and Li, Bao-An",
    title = "{Extracting Nuclear Symmetry Energies at High Densities from Observations of Neutron Stars and Gravitational Waves}",
    eprint = "1807.07698",
    archivePrefix = "arXiv",
    primaryClass = "nucl-th",
    doi = "10.1140/epja/i2019-12700-0",
    journal = "Eur. Phys. J. A",
    volume = "55",
    number = "3",
    pages = "39",
    year = "2019"
}

@article{Li:2021thg,
    author = "Li, Bao-An and Cai, Bao-Jun and Xie, Wen-Jie and Zhang, Nai-Bo",
    title = "{Progress in Constraining Nuclear Symmetry Energy Using Neutron Star Observables Since GW170817}",
    eprint = "2105.04629",
    archivePrefix = "arXiv",
    primaryClass = "nucl-th",
    doi = "10.3390/universe7060182",
    journal = "Universe",
    volume = "7",
    number = "6",
    pages = "182",
    year = "2021"
}

@article{Xie:2020kta,
    author = "Xie, Wen-Jie and Li, Bao-An",
    title = "{Bayesian inference of the incompressibility, skewness and kurtosis of nuclear matter from empirical pressures in relativistic heavy-ion collisions}",
    eprint = "2001.03669",
    archivePrefix = "arXiv",
    primaryClass = "nucl-th",
    doi = "10.1088/1361-6471/abd25a",
    journal = "J. Phys. G",
    volume = "48",
    number = "2",
    pages = "025110",
    year = "2021"
}

@article{Li:2026zuq,
    author = "Li, Bao-An and Grundler, Xavier and Xie, Wen-Jie and Zhang, Nai-Bo",
    title = "{Bayesian Inference of fine-features of dense matter EOS from future high-precision data of neutron star radii}",
    eprint = "2608.04967",
    archivePrefix = "arXiv",
    primaryClass = "astro-ph.HE",
    doi = "10.1051/epjconf/202637806009",
    journal = "EPJ Web Conf.",
    volume = "378",
    pages = "06009",
    year = "2026"
}

@article{Li:2025tku,
    author = "Li, Bao-An and Grundler, Xavier and Xie, Wen-Jie and Zhang, Nai-Bo",
    title = "{Bayesian Inference of Hybrid Star Properties from Future High-precision Measurements of Their Radii}",
    eprint = "2505.00194",
    archivePrefix = "arXiv",
    primaryClass = "astro-ph.HE",
    doi = "10.3847/1538-4357/ae38c0",
    journal = "Astrophys. J.",
    volume = "998",
    number = "2",
    pages = "262",
    year = "2026"
}

@article{Grundler:2025mcz,
    author = "Grundler, Xavier and Li, Bao-An",
    title = "{Bayesian quantification of observability and equation of state of twin stars}",
    eprint = "2506.13677",
    archivePrefix = "arXiv",
    primaryClass = "astro-ph.HE",
    doi = "10.1103/hsd4-j54y",
    journal = "Phys. Rev. D",
    volume = "112",
    number = "10",
    pages = "103012",
    year = "2025"
}

@article{Li:2026ult,
    author = "Li, Bao-An",
    title = "{Universal EOS-Radius Inverse Mappings Govern Precision-Dependent Inference of the Neutron Star Equation of State}",
    eprint = "2606.28183",
    archivePrefix = "arXiv",
    primaryClass = "nucl-th",
    Journal=" to be published",
    month = "6",
    year = "2026"
}

@article{abbott2018gw170817,
  title={GW170817: Measurements of neutron star radii and equation of state},
  author={Abbott, Benjamin P and Abbott, Richard and Abbott, TD and Acernese, F and Ackley, K and Adams, C and Adams, T and Addesso, P and Adhikari, Rana X and Adya, Vaishali B and others},
  journal={Physical review letters},
  volume={121},
  number={16},
  pages={161101},
  year={2018},
  publisher={APS},
  doi = {10.1103/PhysRevLett.121.161101}
}

@article{xie2020bayesian,
  title={Bayesian inference of the symmetry energy of superdense neutron-rich matter from future radius measurements of massive neutron stars},
  author={Xie, Wen-Jie and Li, Bao-An},
  journal={The Astrophysical Journal},
  volume={899},
  number={1},
  pages={4},
  year={2020},
  publisher={IOP Publishing},
  doi={10.3847/1538-4357/aba271}
}

@article{zhang2018combined,
  title={Combined constraints on the equation of state of dense neutron-rich matter from terrestrial nuclear experiments and observations of neutron stars},
  author={Zhang, Nai-Bo and Li, Bao-An and Xu, Jun},
  journal={The Astrophysical Journal},
  volume={859},
  number={2},
  pages={90},
  year={2018},
  publisher={IOP Publishing},
  doi = {10.3847/1538-4357/aac027}
}

@article{seidov,
    author = {{Seidov}, Z.~F.},
    title = {The Stability of a Star with a Phase Change in General Relativity Theory},
    journal = {Soviet Astronomy},
    year = 1971,
    month = oct,
    volume = {15},
    pages = {347},
    adsurl = {https://ui.adsabs.harvard.edu/abs/1971SvA....15..347S}
}

@article{xie2019bayesian,
  title={Bayesian inference of high-density nuclear symmetry energy from radii of canonical neutron stars},
  author={Xie, Wen-Jie and Li, Bao-An},
  journal={The Astrophysical Journal},
  volume={883},
  number={2},
  pages={174},
  year={2019},
  publisher={IOP Publishing},
  doi = {10.3847/1538-4357/ab3f37}
}

@article{xie2021bayesian,
    author = "Xie, Wen-Jie and Li, Bao-An",
    title = "{Bayesian inference of the dense-matter equation of state encapsulating a first-order hadron-quark phase transition from observables of canonical neutron stars}",
    eprint = "2009.13653",
    archivePrefix = "arXiv",
    primaryClass = "nucl-th",
    doi = "10.1103/PhysRevC.103.035802",
    journal = "Phys. Rev. C",
    volume = "103",
    number = "3",
    pages = "035802",
    year = "2021"
}

@article{oppenheimer1939massive,
  title={On massive neutron cores},
  author={Oppenheimer, J Robert and Volkoff, George M},
  journal={Physical Review},
  volume={55},
  number={4},
  pages={374},
  year={1939},
  publisher={APS}
}

@article{tolman1934effect,
  title={Effect of inhomogeneity on cosmological models},
  author={Tolman, Richard C},
  journal={Proceedings of the National Academy of Sciences},
  volume={20},
  number={3},
  pages={169--176},
  year={1934},
  publisher={National Acad Sciences}
}

@article{Lattimer:2000nx,
    author = "Lattimer, J. M. and Prakash, M.",
    title = "{Neutron star structure and the equation of state}",
    eprint = "astro-ph/0002232",
    archivePrefix = "arXiv",
    doi = "10.1086/319702",
    journal = "Astrophys. J.",
    volume = "550",
    pages = "426",
    year = "2001"
}

@article{Chatziioannou:2021tdi,
    author = "Chatziioannou, Katerina",
    title = "{Uncertainty limits on neutron star radius measurements with gravitational waves}",
    eprint = "2108.12368",
    archivePrefix = "arXiv",
    primaryClass = "gr-qc",
    doi = "10.1103/PhysRevD.105.084021",
    journal = "Phys. Rev. D",
    volume = "105",
    number = "8",
    pages = "084021",
    year = "2022"
}

@article{STAR3,
    author = "Abdallah, M. S. and others",
    collaboration = "STAR",
    title = "{Disappearance of partonic collectivity in sNN=3GeV Au+Au collisions at RHIC}",
    eprint = "2108.00908",
    archivePrefix = "arXiv",
    primaryClass = "nucl-ex",
    doi = "10.1016/j.physletb.2022.137003",
    journal = "Phys. Lett. B",
    volume = "827",
    pages = "137003",
    year = "2022"
}

@article{Hild:2009ns,
    author = "Hild, Stefan and Chelkowski, Simon and Freise, Andreas and Franc, Janyce and Morgado, Nazario and Flaminio, Raffaele and DeSalvo, Riccardo",
    title = "{A Xylophone Configuration for a third Generation Gravitational Wave Detector}",
    eprint = "0906.2655",
    archivePrefix = "arXiv",
    primaryClass = "gr-qc",
    doi = "10.1088/0264-9381/27/1/015003",
    journal = "Class. Quant. Grav.",
    volume = "27",
    pages = "015003",
    year = "2010"
}

@article{Sathyaprakash:2012jk,
    author = "Sathyaprakash, B. and others",
    editor = "Hannam, Mark and Sutton, Patrick and Hild, Stefan and van den Broeck, Chris",
    title = "{Scientific Objectives of Einstein Telescope}",
    eprint = "1206.0331",
    archivePrefix = "arXiv",
    primaryClass = "gr-qc",
    doi = "10.1088/0264-9381/29/12/124013",
    journal = "Class. Quant. Grav.",
    volume = "29",
    pages = "124013",
    year = "2012",
    note = "[Erratum: Class. Quant. Grav. 30, 079501 (2013)]"
}

@article{Evans:2021gyd,
    author = "Evans, Matthew and others",
    title = "{A Horizon Study for Cosmic Explorer: Science, Observatories, and Community}",
    eprint = "2109.09882",
    archivePrefix = "arXiv",
    primaryClass = "astro-ph.IM",
    reportNumber = "CE-P2100003-v7, Cosmic Explorer technical report CE-P2100003-v6",
    month = "9",
    journal=" Report",
    year = "2021"
}

@article{Pacilio:2021jmq,
    author = "Pacilio, Costantino and Maselli, Andrea and Fasano, Margherita and Pani, Paolo",
    title = "{Ranking Love Numbers for the Neutron Star Equation of State: The Need for Third-Generation Detectors}",
    eprint = "2104.10035",
    archivePrefix = "arXiv",
    primaryClass = "gr-qc",
    doi = "10.1103/PhysRevLett.128.101101",
    journal = "Phys. Rev. Lett.",
    volume = "128",
    number = "10",
    pages = "101101",
    year = "2022"
}

@article{Finstad:2022oni,
    author = "Finstad, Daniel and White, Laurel V. and Brown, Duncan A.",
    title = "{Prospects for a Precise Equation of State Measurement from Advanced LIGO and Cosmic Explorer}",
    eprint = "2211.01396",
    archivePrefix = "arXiv",
    primaryClass = "astro-ph.HE",
    doi = "10.3847/1538-4357/acf12f",
    journal = "Astrophys. J.",
    volume = "955",
    number = "1",
    pages = "45",
    year = "2023"
}

@article{Bandopadhyay:2024zrr,
    author = "Bandopadhyay, Ananya and Kacanja, Keisi and Somasundaram, Rahul and Nitz, Alexander H. and Brown, Duncan A.",
    title = "{Measuring neutron star radius with second and third generation gravitational wave detector networks}",
    eprint = "2402.05056",
    archivePrefix = "arXiv",
    primaryClass = "astro-ph.HE",
    reportNumber = "LA-UR-24-21031",
    doi = "10.1088/1361-6382/ad828a",
    journal = "Class. Quant. Grav.",
    volume = "41",
    number = "22",
    pages = "225003",
    year = "2024"
}

@article{LIGOScientific:2020zkf,
    author = "Abbott, R. and others",
    collaboration = "LIGO Scientific, Virgo",
    title = "{GW190814: Gravitational Waves from the Coalescence of a 23 Solar Mass Black Hole with a 2.6 Solar Mass Compact Object}",
    eprint = "2006.12611",
    archivePrefix = "arXiv",
    primaryClass = "astro-ph.HE",
    reportNumber = "LIGO-P190814",
    doi = "10.3847/2041-8213/ab960f",
    journal = "Astrophys. J. Lett.",
    volume = "896",
    number = "2",
    pages = "L44",
    year = "2020"
}

@article{Sedrakian:2022ata,
    author = "Sedrakian, Armen and Li, Jia-Jie and Weber, Fridolin",
    title = "{Heavy baryons in compact stars}",
    eprint = "2212.01086",
    archivePrefix = "arXiv",
    primaryClass = "nucl-th",
    doi = "10.1016/j.ppnp.2023.104041",
    journal = "Prog. Part. Nucl. Phys.",
    volume = "131",
    pages = "104041",
    year = "2023"
}

@article{stableMethod,
    author = {{Bardeen}, James M. and {Thorne}, Kip S. and {Meltzer}, David W.},
    title = {A Catalogue of Methods for Studying the Normal Modes of Radial Pulsation of General-Relativistic Stellar Models},
    journal = {Astrophysical Journal},
    year = 1966,
    month = aug,
    volume = {145},
    pages = {505},
    doi = {10.1086/148791},
    adsurl = {https://ui.adsabs.harvard.edu/abs/1966ApJ...145..505B}
}

@article{Alford_2017,
    doi = {10.3847/1538-4357/aa8509},
    url = {https://dx.doi.org/10.3847/1538-4357/aa8509},
    year = {2017},
    month = {sep},
    publisher = {The American Astronomical Society},
    volume = {847},
    number = {2},
    pages = {109},
    author = {Alford, Mark G. and Harris, Steven P. and Sachdeva, Pratik S.},
    title = {On the Stability of Strange Dwarf Hybrid Stars},
    journal = {The Astrophysical Journal}
}

@article{Essick:2024olf,
    author = "Essick, Reed",
    title = "{Exotic Stable Branches with Efficient TOV Sequences}",
    eprint = "2405.05395",
    archivePrefix = "arXiv",
    primaryClass = "astro-ph.HE",
    doi = "10.3847/2041-8213/ad7399",
    journal = "Astrophys. J. Lett.",
    volume = "973",
    number = "2",
    pages = "L50",
    year = "2024"
}

\clearpage
\end{document}